\documentclass{aa}  

\usepackage{amsmath}
\usepackage{graphicx}
\usepackage{changepage}
\usepackage{listings}
\usepackage{siunitx}
\usepackage{natbib}
\usepackage{hyperref}
\usepackage{multirow}
\usepackage{float}
\usepackage{placeins}
\usepackage{pdflscape}  
\usepackage{longtable}  
\usepackage{lastpage}
\usepackage{txfonts}
\newcommand{\err}[1]{\text{\tiny (\(\pm #1\))}}

\usepackage{xcolor}

\begin{document}

   \title{Hydrodynamical mass-loss rates for very massive stars}

   \subtitle{II. New theoretical mass-loss predictions at solar metallicity ($Z= 0.02$)}

\author{{Gautham N. Sabhahit\inst{\ref{AOP}}$^{,\star}$}
    \and 
       {Jorick S. Vink\inst{\ref{AOP}}}
    \and
      {Andreas\,A.\,C. Sander\inst{\ref{ARI}\mathrm{,}\ref{IWR}}}
    }

\institute{
   {Armagh Observatory and Planetarium, College Hill, Armagh BT61 9DG, N. Ireland\label{AOP}}
   \and
   {Zentrum f{\"u}r Astronomie der Universit{\"a}t Heidelberg, Astronomisches Rechen-Institut, M{\"o}nchhofstr. 12-14, 69120 Heidelberg, Germany\label{ARI}}
   \and
   {Interdisziplin{\"a}res Zentrum f{\"u}r Wissenschaftliches Rechnen, Universit{\"a}t Heidelberg, Im Neuenheimer Feld 225, 69120 Heidelberg, Germany\\\label{IWR}}
$^\star$\email{gauthamns96@gmail.com}
             }



  \abstract
{
    The evolutionary pathways and ultimate fates of very massive stars are governed primarily by mass loss through radiatively driven winds. 
    }
    {
    We present a new theoretical mass-loss prescription for (very) massive stars, capturing the complex dependence on the Eddington parameter ($\Gamma_\mathrm{e}$), luminosity, temperature, and hydrogen abundance.
    } 
    {
    We calculated an extensive grid of 178 hydrodynamically self-consistent wind-atmosphere models in non-local thermodynamic equilibrium using the \texttt{PoWR}$^\textsc{hd}$ code, predicting wind properties such as the mass-loss rate and terminal velocity self-consistently. The grid spans masses $M_\star = 16-500\,M_\odot$, luminosities $\log(L_\star/L_\odot) = 5.5-6.8$, inner boundary temperatures $T_\star = 12-50\,\mathrm{kK}$, and hydrogen mass fractions $X = 0.01-0.9$ at a fixed metallicity $Z=0.02$. 
    }
    {
    We confirm the presence of a mass-loss kink in the $\dot{M}$--$\Gamma_\mathrm{e}$ relation across the explored parameter space. The kink marks the transition from a shallow scaling ($\sim$$2$) at low $\Gamma_\mathrm{e}$ for optically thin O-star winds to a steeper scaling ($\sim$$10$) for optically thick winds at high $\Gamma_\mathrm{e}$.  We derive comprehensive fitting relations capturing both the kink behaviour and two bistability jumps arising from iron (Fe) ionisation changes and provide auxiliary relations for implementation into stellar evolutionary calculations. Our prescription correctly reproduces the model-independent transition mass-loss rate in the Arches Cluster, confirming the accuracy of our predicted rates at the O--to--WNh transition. Application of our recipe to the zero-age main sequence provides excellent agreement with recent empirical $\dot{M}$--$\Gamma_\mathrm{e}$ relations obtained for a wide range of temperatures and Eddington parameters.
    }
   {
    We provide a physically motivated, continuous, and empirically anchored mass-loss recipe for (very) massive stars, suitable for stellar evolution calculations in the $20-500\,M_\odot$ range.
    }
    {}


   \maketitle

\section{Introduction}
\label{sec: Introduction}

Very massive stars (VMSs), defined as stars born with masses $M_\mathrm{init} \gtrsim 100\,M_\odot$, play a disproportionately crucial role in shaping their environments despite their scarcity and short lifetimes. In the Local Universe, VMSs are found in young massive clusters such as the Arches and Quintuplet clusters near the Galactic centre and R136 in the Large Magellanic Cloud \citep{Martins2008, Crowther2010, Vink2015, Schneider2018}.  Their extreme luminosities and strong stellar winds inject enormous amounts of energy and momentum into the interstellar medium, while their nucleosynthetic yields -- particularly of nitrogen (N) -- leave lasting imprints on the chemical evolution of galaxies \citep{Yusof2013, Vink2018, Higgins2023}.
These objects are the dominant sources of ionising radiation and mechanical feedback in their host regions, powering giant ionised hydrogen (\ion{H}{II}) regions and driving large-scale outflows that regulate star formation \citep{Krumholz2009}.

From a spectral morphology standpoint, VMSs can be considered as the hydrogen (H)-rich counterparts of classical Wolf-Rayet (WR) stars. Namely, they show strong WR-type emission features formed in their dense stellar winds, while retaining sufficient H in their atmospheres. These dense winds drive strong mass loss, capable of revealing the receding convective core and bringing H-burning nucleosynthesis-cycled material such as N and helium (He) to the surface \citep{Roy2020}. The strong WR-type emission lines, the enhancement in N alongside non-negligible H, give VMSs their WNh spectral classification \citep{Crowther2010}. The He over H enrichment at the surface in turn affects the electron number density and hence the opacity in the atmosphere, with potential consequences for their wind properties.

Mass loss also fundamentally governs the evolutionary pathways and ultimate fates of VMSs. The amount of mass removed by winds determines whether a VMS retains sufficient mass to undergo a pair-instability supernova \citep{Woos07, Farmer2019, Winch2024} or form a heavy black hole. Unlike lower-mass stars where internal mixing dominates the evolution~\citep{Langer2012}, the physics of VMSs is fundamentally governed by wind mass loss primarily during the main sequence (MS). For example, at sufficiently high metallicity, VMS evolution models suggest that these stars shed a large fraction of their initial mass, converging towards similar final masses, largely independent of their initial mass~\citep[e.g.][]{Belkus2007, Yung2008, Kohler2015, Vink2018, Sabhahit2022}.

Despite its critical importance, mass loss at very high masses remains one of the most uncertain aspects of stellar modelling. Empirical mass-loss rate determinations solely based on recombination diagnostics such as H$\alpha$ suffer from degeneracies between the mass-loss rate, clumping, and velocity structure \citep{Hamann1998, Puls2008}. On the other hand, theoretical predictions have historically relied on simplified treatments of radiative transfer or prescribed velocity laws \citep[][the CAK model]{CAK1975}, which may not capture the complex physics of optically thick winds where multiple scattering becomes relevant for WR stars including the WNh sequence \citep{GH2008, Vink2011, Sander2020b}. 

Fortunately, a particularly valuable mass-loss constraint exists in a specific scenario called the transition mass-loss rate. This concept, introduced by \citet{Vink2012}, exploits the spectral transition between O-type stars with optically thin winds and WNh stars with optically thick winds and applies to the transition Of/WNh objects. At this transition point, where the wind optical depth reaches roughly unity, a simple relationship connects the mass-loss rate, terminal velocity, and luminosity: $\eta \equiv \dot{M}\varv_\infty/(L_\star/c) \approx 0.6$. Namely, the wind efficiency also crosses order unity. Because this relation is primarily a luminosity determination, it does not depend on uncertain parameters such as clumping or the ionisation structure that plague mass-loss rate determinations. It provides a robust mass-loss estimate against which wind prescriptions can be tested. Any mass-loss recipe, empirical or theoretical, should reproduce the transition mass-loss rate estimated for Of/WNh objects in young massive clusters, such as the Arches Cluster, using the above wind efficiency arguments. In this sense, the transition mass-loss rate acts as a critical benchmark for assessing the validity of any given mass-loss prescription, and failure to reproduce it would disfavour that prescription.

In this work, we present new hydrodynamically consistent mass-loss predictions for VMSs at solar metallicity ($Z = 0.02$) based on an extensive grid of atmosphere models computed with the hydrodynamical branch of the co-moving frame (CMF) code Potsdam Wolf-Rayet (\texttt{PoWR}). Unlike traditional approaches that prescribe the wind velocity structure a priori, hydrodynamical codes solve the equation of motion self-consistently alongside the radiative transfer, statistical equilibrium, and energy balance equations \citep{GH2005, Sander2017}. This allows us to predict wind parameters such as the mass-loss rate and wind velocity stratification as outputs rather than prescribed inputs, providing a more physically consistent solution that accounts for multiple scattering. 

The CMF approach offers a complementary modelling framework to dynamically consistent Monte Carlo (MC) models. Previous MC studies, for example \citep{Vink2011}, have revealed a pronounced kink in the mass-loss rate as a function of the Eddington parameter (which is proportional to the luminosity-to-mass ratio). Bistable wind solutions have also been found as a function of temperature, arising from changes in the iron (Fe) ionisation balance \citep{Vink99}. The latter phenomenon has now been theoretically predicted by multiple, independent wind codes \citep{Petrov2016, Krticka21, Sabhahit2026a}. We intend to test these features in \texttt{PoWR} across a large parameter range relevant for massive stars and VMSs. 

The paper is organised as follows. In Sect.~\ref{sec: Methods}, we build on our initial study \citep[][hereafter Paper I]{Sabhahit2026a}, and we extend the explored parameter space to include systematic variations in luminosity, mass, temperature, and different He over H enhancement ratios. In Sect.~\ref{sec: results}, we discuss deriving fitting relations that capture the complex dependence of the mass-loss rate on the relevant stellar parameters, reproducing both the mass-loss kink feature at high Eddington parameters and bistability jumps arising from changes in the dominant iron ionisation. We then show that our predictions robustly agree with the transition mass-loss rate in the Arches Cluster and against recent empirical constraints on the mass loss and Eddington parameter relation (Sect.~\ref{sec: Discussion}). The resulting prescription provides a new hydrodynamically consistent mass-loss recipe for massive stars and VMSs for use in stellar evolution calculations.

\section{Hydrodynamically consistent atmosphere model calculations}
\label{sec: Methods}

\begin{table*}[!t]
\centering
\caption{Model sequences used in this work.}
\renewcommand{\arraystretch}{1.15}
\begin{tabular}{llllll}
\hline\hline
Sequence & Fixed Parameters & Varying Parameter & Range & Source & \# of Models \\
\hline
\multicolumn{6}{c}{\textbf{Paper I grid}} \\
\hline
1 & $\log(L_\star/L_\odot) = 6.0$, $X=0.7$ & $M_\star/M_\odot$, $T_\star$ [kK] & $37-115$, $12-50$ & Paper I & 92 \\
\hline
\multicolumn{6}{c}{\textbf{Extension 1: Varying mass at fixed luminosity and H mass fraction}} \\
\hline
2 & $\log(L_\star/L_\odot) = 5.5$, $X=0.7$, $T_\star = 35\,\mathrm{kK}$ & $M_\star/M_\odot$ & $16-35$  & This work & 8 \\
3 & $\log(L_\star/L_\odot) = 6.5$, $X=0.7$, $T_\star = 35\,\mathrm{kK}$ & $M_\star/M_\odot$ & $130-260$ & This work & 10 \\
4 & $\log(L_\star/L_\odot) = 6.8$, $X=0.7$, $T_\star = 35\,\mathrm{kK}$  & $M_\star/M_\odot$ & $250-500$ & This work & 9 \\
5 & $\log(L_\star/L_\odot) = 6.0$, $X=0.5$, $T_\star = 35\,\mathrm{kK}$  & $M_\star/M_\odot$ & $40-90$ & This work & 9 \\
6 & $\log(L_\star/L_\odot) = 6.0$, $X=0.3$, $T_\star = 35\,\mathrm{kK}$  & $M_\star/M_\odot$ & $39-75$ & This work & 7\\
\hline
\multicolumn{6}{c}{\textbf{Extension 2: Varying luminosity and H mass fraction at fixed mass}} \\
\hline
7 & $M_\star/M_\odot = 75$, $X=0.7$, $T_\star = 35\,\mathrm{kK}$  & $\log(L_\star/L_\odot)$ & $5.85-6.25$ & This work & 6 \\
8 & $M_\star/M_\odot = 200$, $X=0.7$, $T_\star = 35\,\mathrm{kK}$ & $\log(L_\star/L_\odot)$ & $6.4-6.65$ & This work & 5 \\
9 & $\log(L_\star/L_\odot) = 6.0$, $M_\star/M_\odot = 50$, $T_\star = 35\,\mathrm{kK}$  & $X$ & $0.01-0.9$ & This work & 7 \\
10 & $\log(L_\star/L_\odot) = 6.0$, $M_\star/M_\odot = 60$, $T_\star = 35\,\mathrm{kK}$  & $X$ & $0.1-0.9$ & This work & 6 \\
\hline
\multicolumn{6}{c}{\textbf{Extension 3: Varying temperature at fixed luminosity and mass}} \\
\hline
11 & $\log(L_\star/L_\odot) = 5.5$, $M_\star/M_\odot = 20$, $X=0.7$  & $T_\star$ [kK] & $23-40$ & This work & 4 \\
12 & $\log(L_\star/L_\odot) = 5.5$, $M_\star/M_\odot = 40$, $X=0.7$ & $T_\star$ [kK] & $15-40$ & This work & 9 \\
13 & $\log(L_\star/L_\odot) = 6.5$, $M_\star/M_\odot = 200$, $X=0.7$  & $T_\star$ [kK] & $20-45$ & This work & 6 \\
\hline
\multicolumn{5}{r}{\textbf{Total:}} & \textbf{92+86 = 178} \\
\hline
\end{tabular}
\label{tab: model_sequences}
\end{table*}

The present work is an extension of Paper I exploring the impact of a wider parameter space on VMS wind properties. As such, the basic setup of the code and the input scheme remain very similar to those in Paper I. Below, we briefly describe our hydrodynamical approach to atmosphere modelling and the resultant predictive power achievable by such modelling, as well as outline the parameter extension explored in this work.

We utilised the hydrodynamical branch of the non-local thermodynamic
equilibrium stellar atmospheric code \texttt{PoWR} \citep{Grafener2002, HG2003, GH2005, GH2008, Sander2015diss, Sander2015, Sander2017} to build our grid of atmosphere models. The fundamental advance of the \texttt{PoWR}$^\textsc{hd}$ models over the base code is the self-consistent solution of the stationary hydrodynamic equation of motion, in addition to the iterations over radiative transfer, statistical and thermal equilibrium, and the equation of continuity typically performed in standard hot-star atmosphere codes. In other words, the code integrates wind hydrodynamics into the overall iterative procedure of atmosphere modelling. The solution is self-consistent in that the mass-loss rate and the wind stratification are calculated and predicted from the net acceleration due to gravity, gradients of gas, turbulent, and radiation pressure rather than being prescribed as in typical empirical atmosphere analyses (e.g. a $\beta$--type velocity law). Further details of the code setup are provided in the Methods section of Paper I. Below, we briefly summarise the main results of Paper I and outline our strategy for extending the model grid.

The Paper I grid comprised hydrodynamically consistent stellar atmosphere models computed at a fixed luminosity of $\log(L_\star/L_\odot)=6.0$, with an H mass fraction $X=0.7$ and metallicity $Z=0.02$. The He mass fraction  was therefore fixed at $Y=1-X-Z = 0.28$, i.e. no enhancement in He over H. Within this framework, we primarily varied the stellar mass, $M_\star$, and the effective temperature, $T_\star$, which was defined at the inner boundary radius $R_\star$, where the Rosseland continuum optical depth reached $\tau_{\mathrm{R,cont}} = 20$. The outer boundary of the models was set at $r_{\mathrm{out}} = 1000\,R_\star$.

At fixed luminosity and composition, varying the stellar mass effectively probed different Eddington parameters. The grid spanned a broad range in classical Eddington parameters, $\Gamma_\mathrm{e} \approx 0.2-0.75$, and stellar temperatures, $T_\star = 12-50\,\mathrm{kK}$. The classical Eddington parameter is defined as the ratio of Thomson acceleration to gravity,
\begin{equation}
\Gamma_\mathrm{e} = \frac{a_\mathrm{thom}}{GM_\star/r^2} \approx \frac{\sigma_\mathrm{e} L_\star}{4 \pi G c M_\star},
\label{eq: eddington_parameter_classical}
\end{equation}
where increasing the mass strengthens gravity and lowers $\Gamma_\mathrm{e}$, and vice versa. 

In the limiting case of complete ionisation of H and He at sufficiently hot temperatures, the electron scattering opacity is given by $\sigma_\mathrm{e} \approx 0.02\cdot(1+X)\,\,\mathrm{m^2\,kg^{-1}}$. The calculation of $\Gamma_\mathrm{e}$ then becomes trivial. Under the assumption of complete ionisation, $\sigma_\mathrm{e}$ is approximately radially constant throughout the atmosphere, making $\Gamma_\mathrm{e}$ a robust and radially constant measure of proximity to the Eddington limit. 

However, if H or He are not completely ionised, then the number of free electrons decreases, lowering $\sigma_\mathrm{e}$. The radial constancy of $\Gamma_\mathrm{e}$ breaks down at cooler temperatures where the approximation of complete ionisation no longer holds, and $\Gamma_\mathrm{e}$ shows a weak outward-decreasing dependence with radius. The $\Gamma_\mathrm{e}$ reported in this work is therefore the radius-averaged value in the sub-critical region\footnote{The critical point in our models occurs where the flow velocity equals the isothermal sound speed corrected for the turbulent velocity, see \citet{Sander2017}.}. 

In Paper I, we identify two distinct behaviours in the predicted mass-loss rates. First, as the stellar mass decreased (or equivalently, as the Eddington parameter increased), the models exhibited a kink-like feature in the mass-loss versus mass relation. We therefore provide an independent validation of the existence of the mass-loss kink, which was first reported using the MC technique in \citet{Vink2011}. The second feature concerns the abrupt switch in wind solutions at specific temperatures, corresponding to a change in the dominant iron ion responsible for driving the wind resulting in bistable winds \citep{Vink99, Petrov2016, Krticka21, VS2021}. 

\begin{figure*}
    \includegraphics[width = \textwidth]{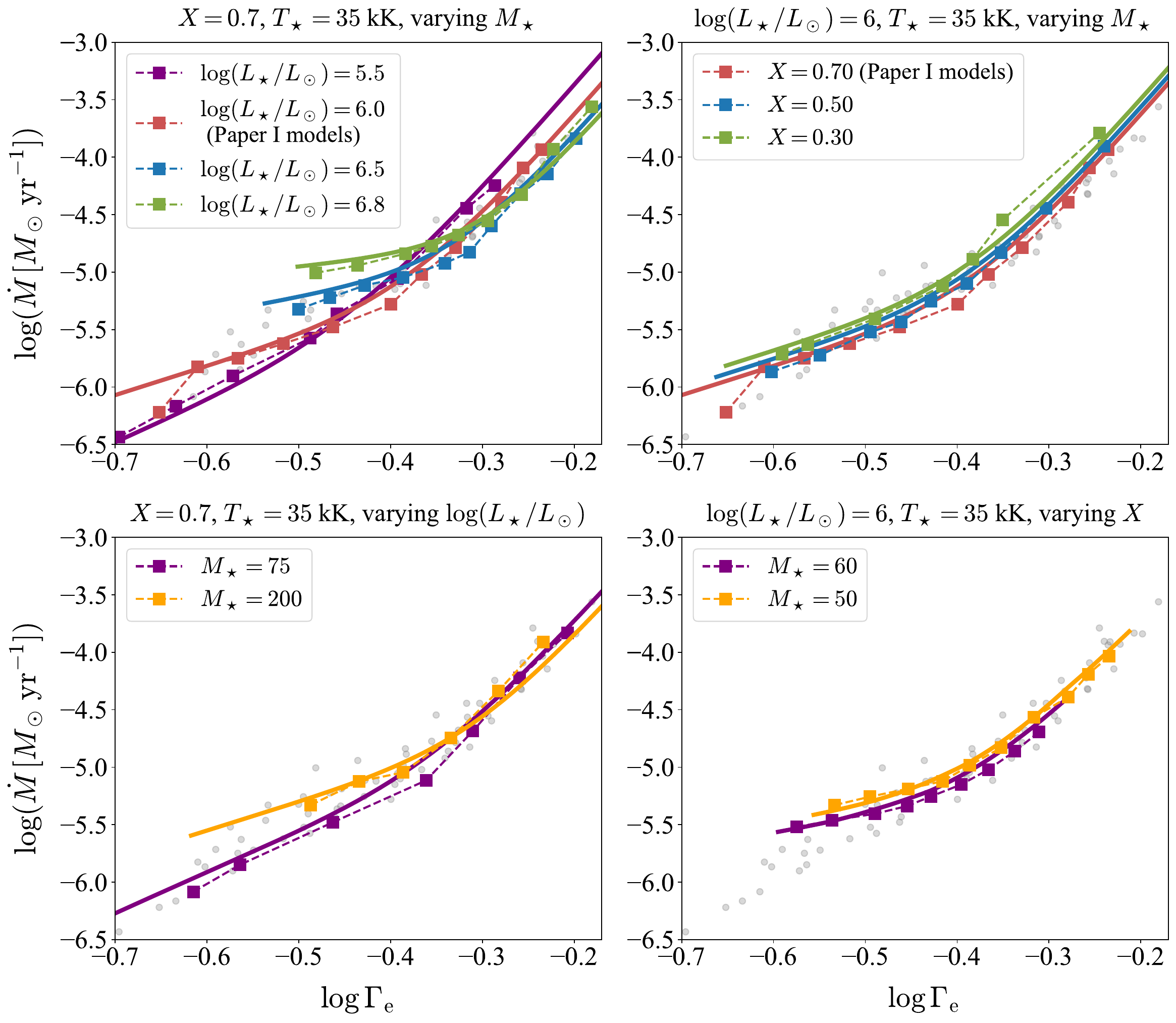}
    \caption{Predicted mass-loss rates from \texttt{PoWR}$^\textsc{hd}$ models as a function of the classical Eddington parameter, $\Gamma_\mathrm{e}$. We examine the effects of varying $L_\star$ and $X$ while keeping the inner boundary temperature fixed at $T_\star = 35\,\mathrm{kK}$ for all the model sequences shown. (\textit{Top:}) Model sequences in which the stellar mass, $M_\star$, is varied at fixed values of $L_\star$ (left panel) and $X$ (right panel). (\textit{Bottom:}) Model sequences in which $L_\star$ (left panel) and $X$ (right panel) are explicitly varied at fixed mass to assess their impact on the mass-loss rate. The coloured symbols connected by dashed lines show the mass-loss rates predicted by our models, while the solid coloured lines indicate the corresponding best-fit relations described in Sect.~3.3. The grey dots mark all models in the grid with $T_\star = 35\,\mathrm{kK}$.}
    \label{fig: mdot_gammae_all}
\end{figure*}

The present work extends the Paper I grid (Seq. 1) in three ways, summarised in Table~\ref{tab: model_sequences}. First, we tested whether the mass-loss kink predicted for a fixed $\log(L_\star/L_\odot)=6.0$ and $X=0.7$ also persists at different luminosities and H mass fractions. A different value of luminosity or H mass fraction was held fixed while the stellar mass was varied. This forms the first extension set of models (Seq. 2-6). Second, we explicitly varied luminosity and the H mass fraction while holding mass and temperature constant, allowing us to probe different combinations of terms in the $\Gamma_\mathrm{e}$ parameter (Seq. 7-10). For both these extension model sets, the inner boundary temperature was held fixed at $T_\star = 35\,\mathrm{kK}$. Third, we explored whether the predicted bistable nature of the winds occurs at different luminosities (Seq. 11-13). Table~\ref{tab: model_sequences} summarises the fixed parameters, the variable parameter, the relevant range, and number of models. 

In total, the overall grid comprises 178 hydrodynamically consistent models, of which 92 were presented in Paper~I, with 86 additional models introduced in the present work. The full set of input parameters and wind-related outputs are provided in Table~\ref{tab: wind_properties}.

All other inputs in this work were fixed to the values adopted in Paper I. This includes the total metallicity, held constant at $Z = 0.02$, with individual metal mass fractions distributed according to solar-scaled abundances from \citet{GS98}. Although the total solar $Z$ has been down-revised in recent years~\citep[e.g.][]{Asplund2009}, the changes to iron have been modest, with most of the decrease occurring in oxygen. Because these massive-star winds are mainly driven by iron~\citep{Vink99, Sabhahit2026a}, there should be a negligible effect on our mass-loss predictions when using recent solar-scaled abundances\footnote{We tested an extreme case scenario where the oxygen mass fraction was set to 0 and the mass-loss rates only decreased by 0.03 dex.}.  The ion line and level lists remain unchanged and are described in detail in the appendix of Paper I.

Wind clumping and turbulence in the atmosphere also take the exact same form as in Paper I. For clumping, we incorporated the micro-clumping formalism with an outward-increasing stratification. The clumping smoothly goes from a smooth wind at the base ($D_\mathrm{cl}=1$) to a clumping factor of $D_\mathrm{cl}=10$ in the outer wind, with the onset of clumping at an optical depth of $\tau_\mathrm{cl}=0.1$. This prescription is motivated by multi-wavelength spectral fits to archetypal VMSs in the Tarantula Nebula, in particular R136a1 (WN5h) and the R144 system, a spectroscopic binary composed of two WNh stars, using the same hydro-version of \texttt{PoWR} \citep{Sabhahit_VMS2025}.

Turbulent pressure was included by adopting a radially constant turbulent velocity of $\varv_\mathrm{turb}=70.71\,\mathrm{km\,s^{-1}}$ for the main grid presented in this work. Such high values were motivated by time-dependent 2D simulations of O-star atmospheres, which predict $\Gamma$--dependent turbulence of $\sim$$10-200\,\mathrm{km\,s^{-1}}$ arising from the hot iron opacity bump \citep{Debnath2024, Moens2025}. The impact of different $\varv_\mathrm{turb}$ values on the predicted mass-loss rates is discussed in Appendix~\ref{appendix: vturb_fits} with an additional set of 44 models. 

\section{Mass-loss predictions for VMSs}
\label{sec: results}

\begin{figure*}
    \includegraphics[width = \textwidth]{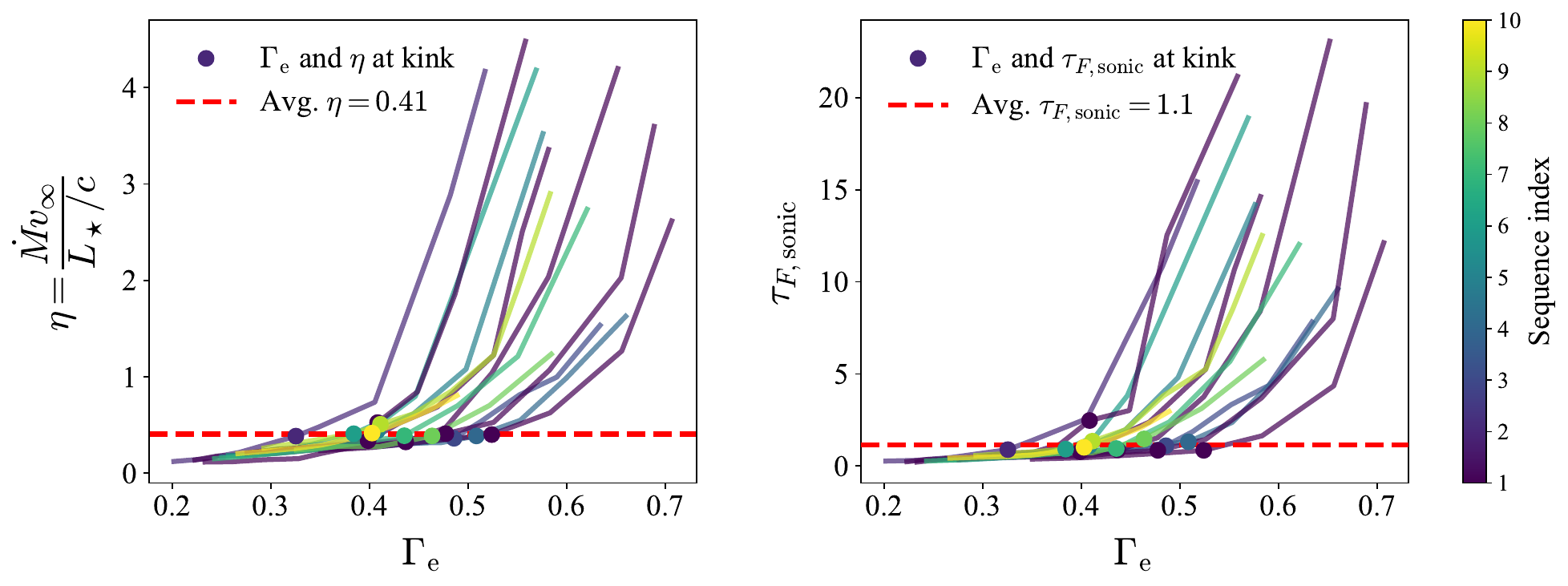}
    \caption{Wind efficiency parameter, $\eta$, and wind optical depth, $\tau_{F,\mathrm{sonic}}$, as a function of the classical Eddington parameter, $\Gamma_\mathrm{e}$. All sequences from Fig.~\ref{fig: mdot_gammae_all} are shown, together with additional model sequences from Paper~I showing a kink. The values of $\Gamma_\mathrm{e}$, $\eta$, and $\tau_{F,\mathrm{sonic}}$ at the location of the kink are indicated by coloured symbols, while the average values, $\eta \approx 0.4$ and $\tau_{F,\mathrm{sonic}} \approx 1$, are shown by dashed red lines. Both the symbols and the solid lines are coloured according to their sequence index, following Table~\ref{tab: model_sequences}.}
    \label{fig: eta_tau}
\end{figure*}

Paper I predicted a kink in the mass-loss rate as a function of stellar mass and, consequently, in the $\dot{M}$--$\Gamma_\mathrm{e}$ relation. The $\dot{M}$--$\Gamma_\mathrm{e}$ power-law slope transitioned from $\sim$$2$ to $\sim$$10$, with the location of the kink occurring at a mass of $M_\mathrm{ref} \approx 60\,M_\odot$, which corresponded to $\Gamma_\mathrm{e}\approx 0.43$. 

At the location of the kink, the models exhibit several characteristic properties. First, the flux-weighted mean wind optical depth at the sonic point crosses unity, while the wind efficiency parameter, which measures the average momentum transfer from radiation to the wind, reaches $\eta \approx 0.45$. Second, the predicted synthetic H$\beta$ profiles display a transitional P Cygni morphology, intermediate between pure absorption and pure emission. Furthermore, models below the kink show absorption-dominated profiles, whereas those above the kink are dominated by emission. These predictions qualitatively agree with the observed spectral transition from absorption-dominated O stars to the emission-dominated spectra characteristic of WNh stars typically found in young-enough, massive clusters. The transitional Of/WNh stars within such clusters indeed exhibit H$\beta$ profiles in P Cygni  \citep[see e.g.][]{CW2011, Best2014}, in agreement with our predictions. Finally, the mass-loss rate at the predicted kink, $\log(\dot{M}\,[M_\odot\,\mathrm{yr}^{-1}]) = -5.22$, agrees to within 0.1 dex with the mass-loss rates inferred for transition Of/WNh stars in the Arches Cluster, both from empirical analyses \citep{Martins2008} and from the concept of the transition mass-loss rate \citep{Vink2012}.

In the present work, we extended the explored parameter space to include variations in $L_\star$ and $X$, and tested whether the wind mass-loss kink persists across this broader parameter domain. Both parameters directly affect the Eddington parameter: the luminosity through the radiative flux, the H abundance through its influence on the number of free electrons, and hence the electron-scattering opacity. We also investigated the second key feature identified in the Paper~I grid, namely the wind bistability behaviour, and its presence at different luminosities. Finally, we present best-fit mass-loss relations that predict absolute mass-loss rates at $Z = 0.02$ over a wide region of parameter space, based on our \texttt{PoWR}$^\textsc{hd}$ models.

\subsection{Wind mass-loss kink}
\label{sec: kink}

Figure~\ref{fig: mdot_gammae_all} summarises the results of our extended $L_\star$ and $X$ grid, showing Sequences~2–10 from Table~\ref{tab: model_sequences}. Without delving into the details of the individual sequences in the first instance, a clear trend is immediately evident: the predicted $\dot{M}$–$\Gamma_\mathrm{e}$ relation exhibits a kink across the entire $L_\star$–$X$ parameter space explored. The mass-loss rate is fundamentally governed by the Eddington parameter, with distinct scaling behaviour in the shallow, low--$\Gamma_\mathrm{e}$ regime and the steeper, high--$\Gamma_\mathrm{e}$ regime, producing the overall kink behaviour.

A similar shallow-to-steep transition was described in Paper~I, where the mass-loss rate was expressed as a power-law function of the stellar mass $M_\star$, smoothly connecting low- and high-mass regimes. For a fixed luminosity and surface composition, these correspond to high- and low--$\Gamma_\mathrm{e}$ regimes, respectively, but the effects of varying $L_\star$ and He enhancement at the surface were not explicitly tested. Here, by varying $M_\star$, $L_\star$, and $X$, we find that, to first order, it is the combination of these quantities entering Eq.~\eqref{eq: eddington_parameter_classical} giving $\Gamma_\mathrm{e}$ that sets the mass-loss rate. That $\Gamma_\mathrm{e}$ so fundamentally controls the mass-loss behaviour is in excellent agreement with previous theoretical predictions using different codes and methods \citep{GH2008, Vink2011}, as well as with empirical studies of massive-star winds across the relevant parameter space \citep{Best2014, Brands22, Pauli2025}.

With kinks predicted ubiquitously across the explored parameter space, we expressed our power-law scaling and the transition from shallow to steep regimes as a function of $\Gamma_\mathrm{e}$. To describe this transition, we adopted a smooth connection between the two regimes using a log-sum-exponential formulation,
\begin{equation}
\begin{split}
\log \dot{M} \propto \log \Big[
10^{\,f_\mathrm{low}(\Gamma_\mathrm{e}/\Gamma_\mathrm{e,ref})}
+ 10^{\,f_\mathrm{high}(\Gamma_\mathrm{e}/\Gamma_\mathrm{e,ref})}
\Big].
\label{eq: mdot_kink}
\end{split}
\end{equation}
Depending on whether $\Gamma_\mathrm{e}$ is smaller or larger compared to the reference value at the kink, $\Gamma_\mathrm{e,ref}$, one of the two terms dominates and sets the scaling. For example, if for a star $\Gamma_\mathrm{e} \ll \Gamma_\mathrm{e,ref}$,  the first term inside the log function dominates. The entire expression then simplifies to $\log\dot{M} \propto f_\mathrm{low}(\Gamma_\mathrm{e}/\Gamma_\mathrm{e,ref})$, which sets the shallow, low--$\Gamma_\mathrm{e}$ slope relevant for O stars. Vice versa, if $\Gamma_\mathrm{e} \gg \Gamma_\mathrm{e,ref}$, the second term inside the log function dominates and the expression simplifies to $\log\dot{M} \propto f_\mathrm{high}(\Gamma_\mathrm{e}/\Gamma_\mathrm{e,ref})$; that is, the high--$\Gamma_\mathrm{e}$ slope relevant for VMSs dominates.

We also find a spread in the predicted mass-loss rates for a fixed value of $\Gamma_\mathrm{e}$ depending on the value of $L_\star$ or $X$.  Therefore, a mass-loss scaling with $\Gamma_\mathrm{e}$ alone is insufficient to fully capture the complexity of our models. 
All four sub-panels in Fig.~\ref{fig: mdot_gammae_all} reveal a primary scaling with $\Gamma_\mathrm{e}$, but also additional explicit dependencies on $L_\star$ and $X$. This is qualitatively consistent with mass-loss trends predicted by MC models, where combinations of $\Gamma_\mathrm{e}$ and $L_\star$, $\Gamma_\mathrm{e}$ and $M_\star$, or $L_\star$ and $M_\star$ were required to reproduce the results. The spread in the mass-loss rates is also expected from empirical results, where stars spanning a range of luminosities, temperatures, and masses show similar scatter \citep[e.g.][]{Pauli2025}.

The need for such a combination of quantities and the failure of a single $\Gamma_\mathrm{e}$ scaling makes physical sense, as $L_\star$, $M_\star$ and even $X$ have fundamentally different effects on $\dot{M}$~\citep[see tests in][]{Sabhahit_VMS2025}. Higher luminosity increases the radiative force, driving stronger winds. Higher mass on the other hand increases gravity, which must be overcome to drive a wind, thereby suppressing $\dot{M}$. The H mass fraction controls the electron number density, which sets the base electron scattering radiative acceleration. A higher $X$ increases $\Gamma_\mathrm{e}$ in the atmosphere, bringing the star closer to its Eddington limit and thus increasing the predicted $\dot{M}$. Although $L_\star$, $M_\star$, and $X$  together influence the overall $\Gamma_\mathrm{e}$ scaling, there is no reason to expect them to contribute equally such that a simple $\Gamma_\mathrm{e}$ scaling alone would suffice. In other words, our final prescription requires explicit $L_\star$ and $X$ scalings on top of the overall $\Gamma_\mathrm{e}$--dependent kink behaviour described by Eq.~\eqref{eq: mdot_kink}.

Another notable feature concerns the location of the kink in $\Gamma_\mathrm{e}$--space, that is, the value of $\Gamma_\mathrm{e,ref}$. The top and bottom left sub-panels (Seq. $2-4$ and 7–8) indicate a weak dependence of the kink location on luminosity, with the location of the kink shifting to higher $\Gamma_\mathrm{e}$ values as $L_\star$ increases. In contrast, the right sub-panels probing the effects of $X$ (Seq. $5-6$ and $9–10$) show no systematic dependence on $X$. Accordingly, we allowed $\Gamma_\mathrm{e,ref}$ to vary with $L_\star$, and, combined with the weak temperature dependence identified in Paper~I, we adopted
\begin{equation}
\Gamma_\mathrm{e,ref} \propto f(L_\star, T_\star)
\label{eq: gamma_e_ref}.
\end{equation}
The final fit relations, shown as solid coloured lines in Fig.~\ref{fig: mdot_gammae_all}, incorporate these dependencies for both $\dot{M}$ and $\Gamma_\mathrm{e,ref}$ (see Sect.~\ref{sec: fit_relations}).

\begin{figure}
    \includegraphics[width = \columnwidth]{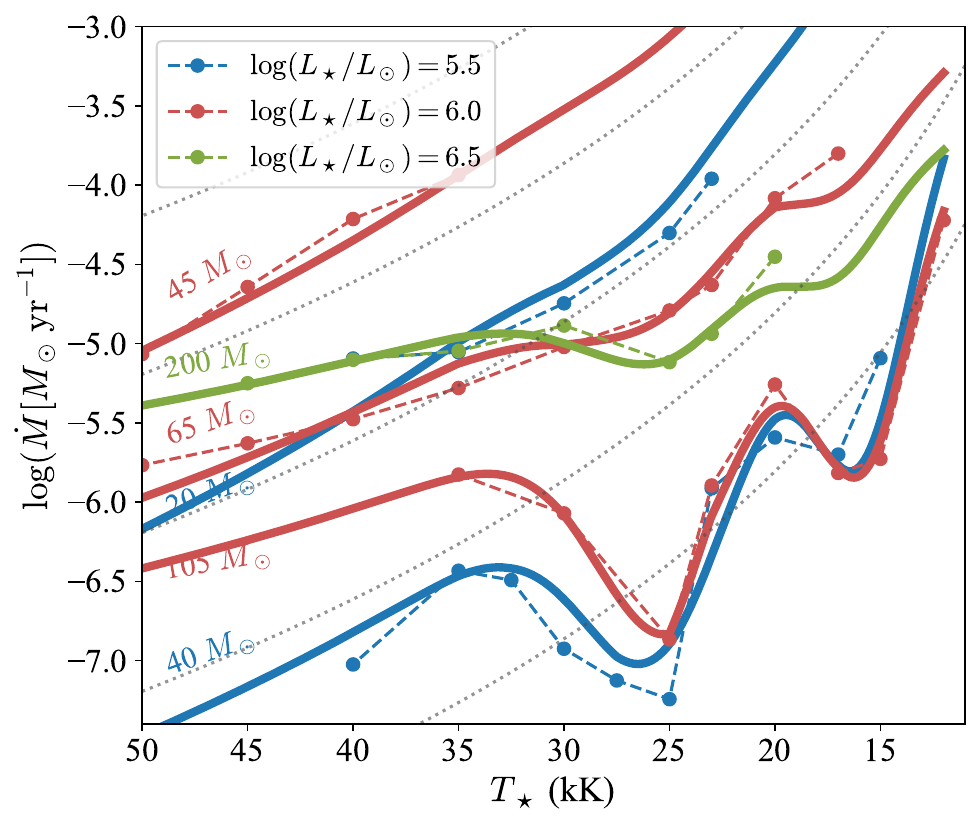}
    \caption{Predicted mass-loss rates from \texttt{PoWR}$^\textsc{hd}$ models as a function of the inner boundary temperature $T_\star$ for a select few model sequences from our grid. All stellar parameters are held fixed, while $T_\star$ is varied. The stellar mass of each sequence is labelled in the plot, while the luminosity is given in the legend. The H mass fraction is fixed at $X=0.7$. The dashed black lines indicate a reference $\dot{M} \propto T_\star^{-6}$ scaling.}
    \label{fig: BSJ}
\end{figure}

Finally, we examined the wind efficiency parameter and the wind optical depth at the location of the kink across our extended model grid. The wind efficiency parameter, $\eta$, is defined as the ratio of the wind momentum to the total radiative momentum available from the star,
\begin{equation}
\eta = \frac{\dot{M} \varv_\infty}{L_\star / c},
\label{eq: eta_wind_efficiency}
\end{equation}
where $\varv_\infty$ is the wind velocity at the outer boundary, also called terminal velocity.
\citet{Vink2011} identified the kink in their MC models as occurring where the single-scattering limit is approximately breached. This was further elaborated by \citet{Vink2012}, who connected the kink to the wind optical depth $\tau_{F,\mathrm{sonic}}$ simultaneously reaching values of order unity. A simple condition is therefore obtained: $\eta \approx 0.6\tau_{F,\mathrm{sonic}}$ and $\tau_{F,\mathrm{sonic}}\approx 1$. This relation could be applied directly to the observed transition objects to infer a model-independent mass-loss rate \citep[see][for more details]{Vink2012}.

In Fig.~\ref{fig: eta_tau}, we show the variation in $\eta$ and $\tau_{F,\mathrm{sonic}}$ as a function of $\Gamma_{\mathrm{e}}$, marking the corresponding values of $\Gamma_{\mathrm{e}}$, $\eta$, and $\tau_{F,\mathrm{sonic}}$ at the kink location for each sequence. Averaging over all sequences, we obtain $\langle \eta \rangle \approx 0.4$ and $\langle \tau_{F,\mathrm{sonic}} \rangle \approx 1$ at the kink. The mean wind efficiency is slightly lower compared to previous estimates, which suggest $\eta \approx 0.6$~\citep{Vink2012}. A similar under-prediction of $\eta$ is found in Paper~I when closely matching the stellar properties of the transition objects in the Arches Cluster, where the wind efficiency at the kink reaches $\eta \approx 0.45$. This discrepancy likely reflects our systematically under-predicted terminal wind velocities compared to those inferred for transition-type stars in the Arches Cluster. The possible under-prediction of $\eta$ at the kink, however, is mostly irrelevant for our final mass-loss recipe, as we parametrised our recipe based on the value of $\Gamma_{\mathrm{e}}$ and $\Gamma_{\mathrm{e,ref}}$.

More importantly, we find that, while the kink spans a significant fraction of the explored $\Gamma_{\mathrm{e}}$ range, both $\eta$ and $\tau_{F,\mathrm{sonic}}$ at the kink occupy a relatively narrow range compared to their overall variation across the grid. For example, $\tau_{F,\mathrm{sonic}}$ varies from $\approx 0.1$ to $\approx 20$ in the whole grid, but the optical depth at the kink across all sequences is confined to a narrow range around $\approx 1$, suggesting that the physics of multiple scattering and the optical thickness of the wind fundamentally sets the location of the kink over $\Gamma_{\mathrm{e}}$. 

An important consequence of both the wind efficiency parameter and wind optical depth crossing order unity at the location of the kink is that the present prescription is entirely consistent with the \citet{Sabhahit2022} $\eta$--framework for VMS mass loss. \citet{Sabhahit2022} implemented a kink formalism with a steep $\dot{M}$--$\Gamma_\mathrm{e}$ above the kink based on the MC models~\citep{Vink2011} and used the $\eta$ criterion to anchor their absolute rates to the transition mass-loss rate observed in the Arches Cluster~\citep{Vink2012}. Here, we instead predict an explicit kink in the $\dot{M}$--$\Gamma_\mathrm{e}$ relation, with shallow and steep slopes for low- and high--$\Gamma_\mathrm{e}$ captured in a single continuous prescription. The two formalisms are qualitatively consistent because both show the kink at the location where the single scattering limit is roughly exceeded. The only difference is that here we explicitly provide the $\dot{M}$ scaling with $X$, and the high--$\Gamma_\mathrm{e}$ slope above the kink quantitatively differs from the MC predictions (see Sect.~\ref{sec: fit_relations}).

\subsection{Bistable winds}
\label{sec: BSJ}

The temperature dependence of stellar wind strength has been debated in the literature for decades, both from theoretical wind models and empirical analyses \citep{Vink99, Bjorklund2021, Verhamme2024}. Part of this dependence is a purely radius effect: at fixed luminosity, lower effective temperatures imply larger stellar radii. Mass removal then becomes easier because the outer layers are more weakly bound, being located farther out in the gravitational potential.

At the same time, cooler temperatures create a mismatch between the stellar flux peak and the dominant Fe driving lines in the ultraviolet (UV). Changes in the ionisation balance of Fe--the primary wind driver at high $Z$--can lead to a switch in the wind solution. This has been most pronounced in distinct `jumps' in the mass-loss rate  identified in MC calculations  when iron recombines~\citep{Vink99, VS2021}. More recently, CMF-based codes probing also higher mass-loss regimes revealed that these jumps can be part of `valleys' where the mass-loss rate is lower due to insufficient Fe driving~\citep{Krticka21, Lefever2025, Sabhahit2026a, Bernini-Peron2026}, although debate persists between wind modelling predictions and empirical analyses of B supergiants~\citep{Vink2026}.

\begin{figure}
    \includegraphics[width = \columnwidth]{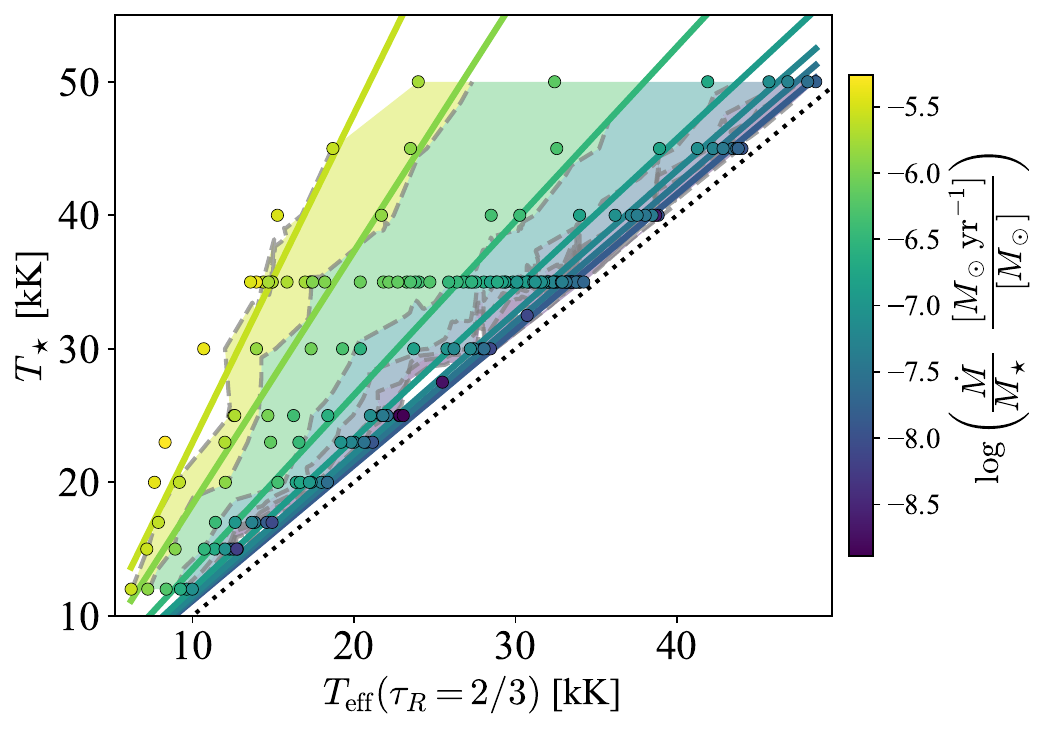}
    \caption{Inner boundary temperature $T_\star$ as a function of the effective temperature at a Rosseland optical depth $\tau_{\mathrm{R}} = 2/3$ as winds become optically thick. The symbols mark our $\texttt{PoWR}^{\textsc{hd}}$ model predictions, colour-coded by the mass-loss rate. Contours of constant $\dot{M}/M_\star$ (in $M_\odot \,\mathrm{yr^{-1}}\,M_\odot^{-1}$) are shown with dashed lines, while the solid lines denote the fits from Eq.~\eqref{eq: Tstar_T23_Mdot_fit}. The dotted black line marks the one-to-one relation.}
    \label{fig: Tstar_Teff}
\end{figure}

\begin{figure}
    \includegraphics[width = \columnwidth]{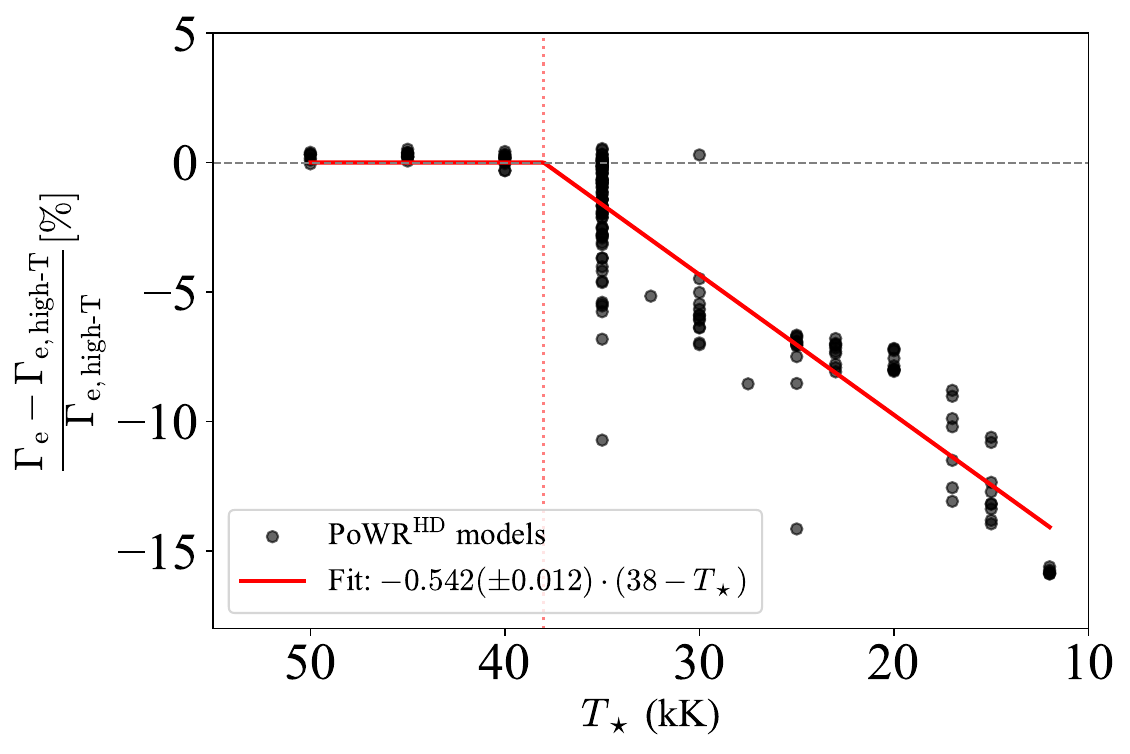}
    \caption{Percentage difference between the $\Gamma_\mathrm{e}$ predicted in the inner wind of our \texttt{PoWR}$^\textsc{hd}$ models (accounting for ionisation effects) and the high-temperature limit, $\Gamma_\mathrm{e,high\text{-}T}$, obtained by assuming complete H and He ionisation (Eq.~\ref{eq: eddington_parameter_classical_high_temp}). The solid red line is our best-fit relation accounting for the growing difference with decreasing $T_\star$. Zero deviation is shown as a dashed black line.}
    \label{fig: Gama_e_fit}
\end{figure}

In Fig.~\ref{fig: BSJ}, we predict two bistability jumps at $T_\star \approx 25\,\mathrm{kK}$ and $T_\star \approx 17\,\mathrm{kK}$, corresponding to ionisation changes from \ion{Fe}{IV} to \ion{Fe}{III} and from \ion{Fe}{III} to \ion{Fe}{II}, respectively. The bistable behaviour is most pronounced at low $\Gamma_\mathrm{e}$ and is present regardless of the luminosity or stellar mass, whereas radius effects dominate at higher $\Gamma_\mathrm{e}$. Our mass-loss fitting relations must therefore incorporate both contributions: a smooth, monotonic, generally increasing mass-loss trend towards larger radii, combined with superposed bistability jumps whose depth varies with $\Gamma_\mathrm{e}$.

\subsection{Mass-loss fitting relations}
\label{sec: fit_relations}

The mass-loss prescription combines a log-sum-exponential scaling in $\Gamma_\mathrm{e}$, explicit scalings with luminosity, H mass fraction and temperature, and two bistability dips to capture the complex behaviour across parameter space. We simultaneously fit all 178 models from Table~\ref{tab: model_sequences} to the following functional form:

\begin{equation}
\begin{split}
\log\dot{M} &= \log\dot{M}_\mathrm{o}
+ \log \Big[
10^{\,f_\mathrm{low}(\Gamma_\mathrm{e})}
+ 10^{\,f_\mathrm{high}(\Gamma_\mathrm{e})}
\Big] \\
&\quad +  f_\mathrm{L} \log\bigg(\dfrac{L_\star}{10^6}\bigg)
+ f_\mathrm{X}\log\bigg(\dfrac{1+X}{1.7}\bigg) + f_\mathrm{T} \log \bigg(\dfrac{T_\star}{T_\mathrm{ref}}\bigg)  \\
&\quad - A_1 \exp\!\Bigg[
k_1 \log\bigg(\dfrac{\Gamma_\mathrm{e}}{\Gamma_\mathrm{e,ref}}\bigg)
\Bigg]
\exp\!\Bigg[
- \left(\dfrac{T_\star - T_1}{\sigma_1}\right)^2
\Bigg] \\
&\quad - A_2 \exp\!\Bigg[
k_2 \log\bigg(\dfrac{\Gamma_\mathrm{e}}{\Gamma_\mathrm{e,ref}}\bigg)
\Bigg]
\exp\!\Bigg[
-\left(\dfrac{T_\star - T_2}{\sigma_2}\right)^2
\Bigg].
\label{eq: mdot_full_fit}
\end{split}
\end{equation}

The inputs are $\Gamma_\mathrm{e}$, $L_\star$ (in $L_\odot$), $T_\star$ (in kK), and $X$ which controls the He over H enhancement as $Z$ is fixed at $0.02$.\footnote{Note that we chose the $1+X$ functional form for the scaling with H mass fraction as it appears naturally in the $\Gamma_\mathrm{e}$ expression under the complete ionisation approximation.} All logarithms are base 10. Mass enters only implicitly via $\Gamma_\mathrm{e}$. The coefficients are
\begin{align*}
&\log\dot{M}_\mathrm{o} = -5.448 \err{0.021}, \\
&f_\mathrm{low}(\Gamma_\mathrm{e}) =
\bigg[
2.240 \err{0.181}
- 2.023 \err{0.431}
\cdot\log\bigg(\dfrac{L_\star}{10^6}\bigg)
\bigg]\cdot
\log\bigg(\dfrac{\Gamma_\mathrm{e}}{\Gamma_\mathrm{e,ref}}\bigg), \\
&f_\mathrm{high}(\Gamma_\mathrm{e}) =
9.243 \err{0.277}\cdot
\log\bigg(\dfrac{\Gamma_\mathrm{e}}{\Gamma_\mathrm{e,ref}}\bigg), \\
&f_\mathrm{L} = 0.762 \err{0.072},
\quad
f_\mathrm{X} = -1.150 \err{0.262}, \\
&f_\mathrm{T} =
-3.857 \err{0.247}
+ 3.772 \err{0.648}
\cdot \log\bigg(\dfrac{L_\star}{10^6}\bigg), \\
&A_1 = 0.367 \err{0.057},
\quad
k_1 = -5.956 \err{0.641}, \\
&A_2 = 0.346 \err{0.074},
\quad
k_2 = -8.019 \err{0.803}, \\
&T_2 = 17\,\mathrm{kK} + 6.095 \err{0.730}
\cdot \log\bigg(\dfrac{\Gamma_\mathrm{e}}{\Gamma_\mathrm{e,ref}}\bigg), \\
&T_1 = 25\,\mathrm{kK},
\quad
\sigma_1 = 5\,\mathrm{kK},
\quad
\sigma_2 = 3\,\mathrm{kK},
\quad
T_\mathrm{ref} = 38\,\mathrm{kK}, \\
&\Gamma_\mathrm{e,ref} =
0.43
+ 0.112 \err{0.012}\cdot\log\bigg(\dfrac{L_\star}{10^6}\bigg)
+ 0.166 \err{0.022}
\left(\dfrac{T_\star}{38} - 1\right).
\end{align*}
Our best-fit relations are shown in Fig.~\ref{fig: mdot_gammae_all} and Fig.~\ref{fig: BSJ}. The relation simultaneously captures both the wind mass-loss kink as a function of $\Gamma_\mathrm{e}$ across varying $L_\star$, $M_\star$, and $X$, and the bistability behaviour with $\Gamma_\mathrm{e}$--dependent jump strengths. At low $\Gamma_\mathrm{e}$, the $\dot{M}$--$\Gamma_\mathrm{e}$ scaling has a shallow slope of $\sim$$2$, although a weak luminosity dependence was required for adequate fits. Above the kink, the slope steepens to $\sim$$10$ and remains roughly constant. The typical root-mean-squared error in the mass-loss fit relation is 0.12 dex.

Implementation in stellar evolution codes faces two practical limitations. The first limitation, already pointed out in Paper I, concerns $T_\star$, defined at a fixed Rosseland continuum optical depth of 20. In our $\texttt{PoWR}^{\textsc{hd}}$ models, $T_\star$ is an input while $T_\mathrm{eff}(\tau_\mathrm{R} = 2/3)$ is an output, dependent on wind strength. For optically thin models at low $\dot{M}$, the difference between the two temperatures is negligible, asymptoting to roughly 1 kK. As $\dot{M}$ and the density scale height increase, the difference between the two temperatures can easily reach $\sim$$10$ kK \citep[see][for more detail]{Smith2004}.

Most structure codes, however, compute surface temperature at $\tau_\mathrm{R} = 2/3$, assuming a static, grey atmosphere. We therefore provide a complementary relation connecting the two temperatures as predicted in our $\texttt{PoWR}^{\textsc{hd}}$ models:
\begin{equation}
T_\star = a + b\cdot T_{\mathrm{eff}}(\tau_{\mathrm{R}} = 2/3),
\label{eq: Tstar_T23_Mdot_fit}
\end{equation}
where
\begin{equation}
\begin{split}
a &= 1 - 0.221 \err{0.059}\; f(\dot M, M_\star), \\
b &= 1 + 0.122 \err{0.004}\; f(\dot M, M_\star),\\
f(\dot M, M)
&=
\frac{\displaystyle \dfrac{\dot M/M_\star}{6.92\times10^{-6}\,M_\odot\,\mathrm{yr}^{-1}/60\,M_\odot}}
{\displaystyle 1
+ 0.040 \err{0.004}\,
\dfrac{\dot M/M_\star}{6.92\times10^{-6}\,M_\odot\,\mathrm{yr}^{-1}/60\,M_\odot}},
\end{split}
\end{equation}
with both temperatures expressed in kilokelvin. Figure~\ref{fig: Tstar_Teff} shows the $T_\star$--$T_{\mathrm{eff}}(\tau_{\mathrm{R}} = 2/3)$ best-fit relation. In the low-$\dot{M}$ limit, the formula reduces to $T_\star \approx 1 + T_{\mathrm{eff}}(\tau_{\mathrm{R}} = 2/3)$, a constant $1\,\mathrm{kK}$ offset in agreement with our models. The typical root-mean-squared error in the $T_\star$--$T_{\mathrm{eff}}(\tau_{\mathrm{R}} = 2/3)$ fit relation is $1\,\mathrm{kK}$.

\begin{figure}
    \includegraphics[width = \columnwidth]{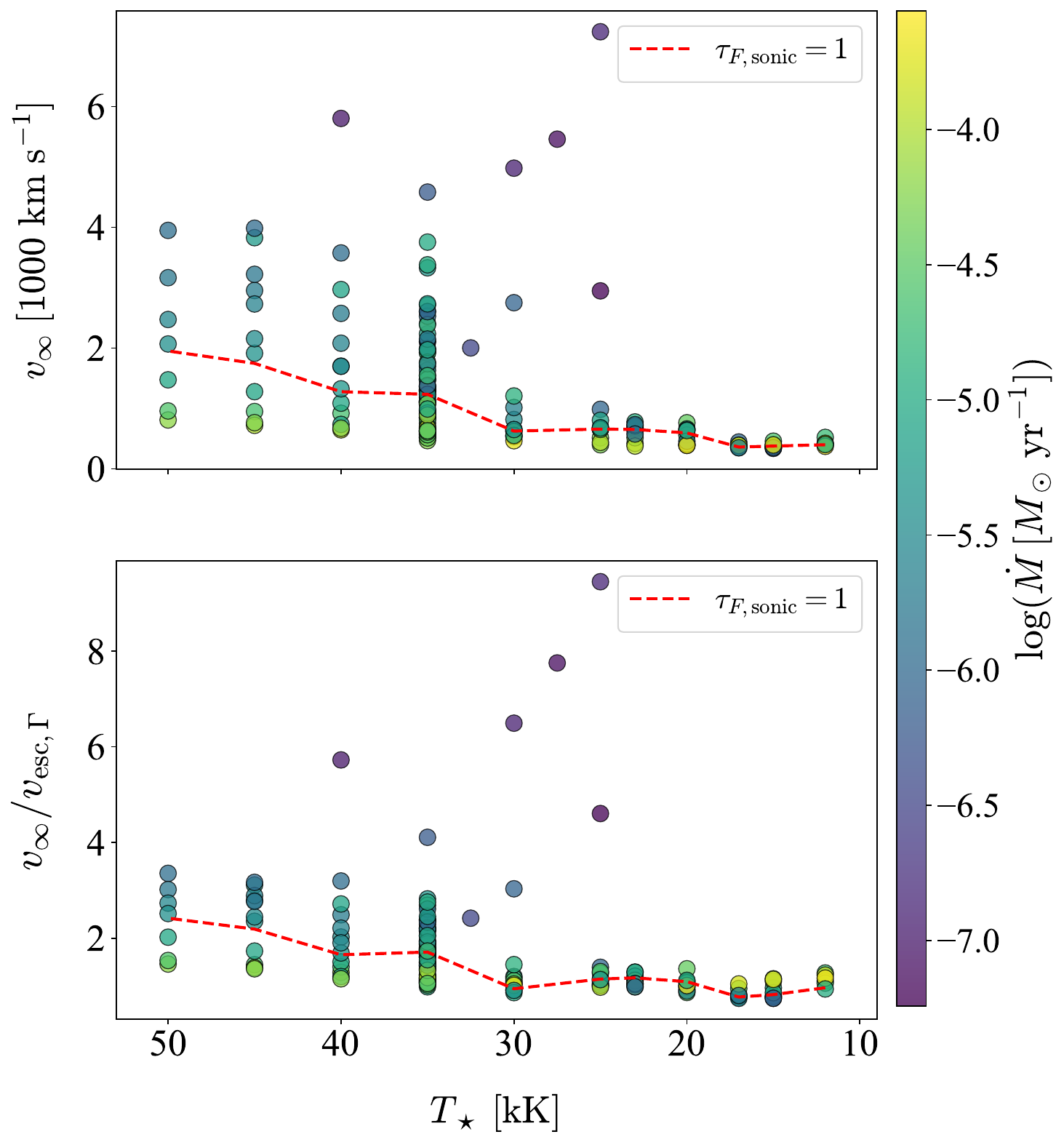}
    \caption{(\textit{Top:}) Terminal velocity. (\textit{Bottom:}) Terminal-to-escape velocity ratio as a function of temperature, $T_\star$, for all models in our grid. Individual symbols are colour-coded according to their mass-loss rate. The dashed red line marks where the flux-mean optical depth at the sonic point, $\tau_{F,\mathrm{sonic}}$, roughly equals unity. The models above the dashed red line indicate $\tau_{F,\mathrm{sonic}} < 1$, and the models below it indicate $\tau_{F,\mathrm{sonic}} > 1$.}
    \label{fig: vfinal_all}
\end{figure}

\begin{figure}
    \includegraphics[width = \columnwidth]{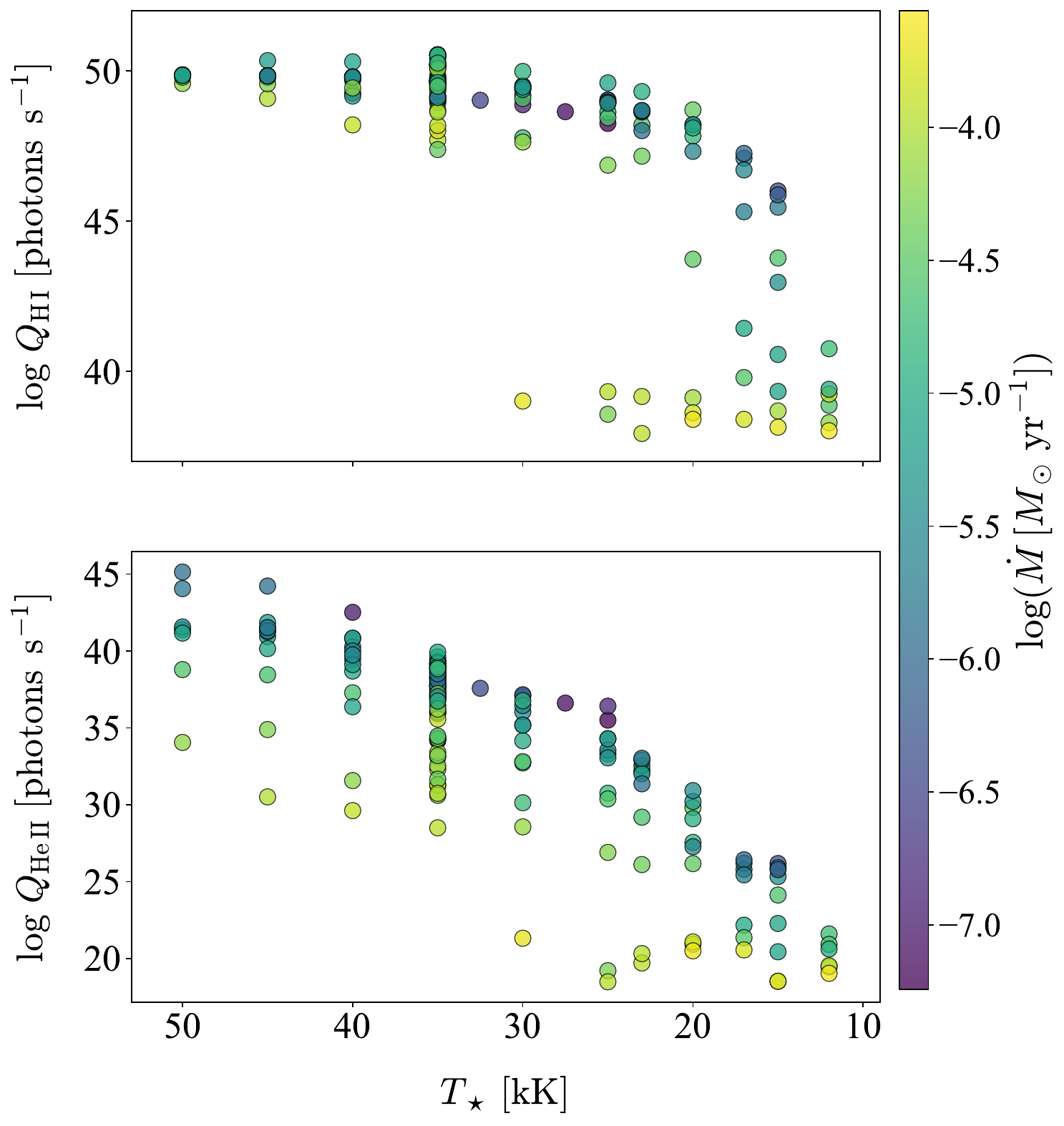}
    \caption{(\textit{Top:}) \ion{H}{I}. (\textit{Bottom:}) \ion{He}{II} ionising flux counts (in log of number of photons per second) as a function of temperature, $T_\star$, for all models in our grid. Individual symbols are colour-coded according to their mass-loss rate.}
    \label{fig: ionising_flux}
\end{figure}

The second issue concerns the calculation of $\Gamma_\mathrm{e}$. At sufficiently high temperatures ($T(r) > 30\,\mathrm{kK}$), Eq.~\eqref{eq: eddington_parameter_classical} reduces to 
\begin{equation}
\Gamma_\mathrm{e,high\text{-}T}= 10^{-4.813} \cdot(1+X) \cdot \dfrac{L_\star/L_\odot}{M_\star/M_\odot},
\label{eq: eddington_parameter_classical_high_temp}
\end{equation}
which is trivially evaluated from structure models. At cooler temperatures, Eq.~\eqref{eq: eddington_parameter_classical_high_temp} overestimates $\Gamma_\mathrm{e}$ by roughly $10-15\%$. Given the steep $\dot{M}$--$\Gamma_\mathrm{e}$ scaling at high $\Gamma_\mathrm{e}$, even $10\%$ over-prediction matters. For example, if Eq.~\eqref{eq: eddington_parameter_classical_high_temp} yields $\Gamma_\mathrm{e} = 0.5$ while the actual value is $\Gamma_\mathrm{e}\approx 0.45$, this produces roughly 0.5 dex over-prediction in $\dot{M}$ along the steep branch of the $\dot{M}$--$\Gamma_\mathrm{e}$ kink.

We therefore provide a temperature-dependent correction term that accounts for the reduction in the $\Gamma_\mathrm{e}$ at cooler temperatures. We used a simple piecewise linear relation for the correction which vanishes for hotter temperatures:
\begin{equation}
\Gamma_\mathrm{e} = \Gamma_\mathrm{e,high\text{-}T} \left(1 - \frac{0.542\err{0.012} \cdot \max(38 - T_\star, 0)}{100}\right),
\label{eq: Gamma_e_fit}
\end{equation}
where $T_\star$ is expressed in kilokelvin. The best-fit relation is shown in Fig.~\ref{fig: Gama_e_fit} and accounts for the growing discrepancy between the \texttt{PoWR}$^\textsc{hd}$ predicted $\Gamma_\mathrm{e}$ and $\Gamma_\mathrm{e,high\text{-}T}$ as $T_\star$ decreases. Above a certain temperature limit, H and He are fully ionised, and the deviation is zero. The typical root-mean-squared error in the $\Gamma_\mathrm{e}$--correction fit is $1.7\%$. While scatter exists in both the $T_\star$--$T_{\mathrm{eff}}(\tau_{\mathrm{R}} = 2/3)$ and $\Gamma_\mathrm{e}$--correction fits, the above-mentioned general trend from our \texttt{PoWR}$^\textsc{hd}$ models is well captured. We note that these are approximate corrections intended for evolutionary implementation, and some scatter is an inherent limitation of any such formula.

Both corrections are relevant for stellar structure and evolution modelling. Given $T_\mathrm{eff}(\tau_\mathrm{R} = 2/3)$ from the model structure, a first approximation to $\Gamma_\mathrm{e}$ from Eq.~\eqref{eq: eddington_parameter_classical_high_temp}, and an initial guess for $T_\star > T_\mathrm{eff}(\tau_\mathrm{R} = 2/3)$, we iterated through Eqs.~\eqref{eq: mdot_full_fit}, \eqref{eq: Tstar_T23_Mdot_fit} and \eqref{eq: Gamma_e_fit} to obtain $\dot{M}$ consistent with the input $T_\mathrm{eff}(\tau_\mathrm{R} = 2/3)$ and the first approximation for $\Gamma_\mathrm{e,high\text{-}T}$.

We also tested varying $\varv_\mathrm{turb}$, and our main result is that $\varv_\mathrm{turb}$ has the strongest effect on the predicted $\dot{M}$ at the coolest temperatures. If VMSs remain hot and compact during most of their evolution (a possibility given their strong winds), then $\varv_\mathrm{turb}$ has a minimal effect on their evolution. However, canonical massive stars that evolve towards cooler supergiants during their MS can be significantly affected depending on the details of turbulence in the atmosphere.

\subsection{Terminal velocities}
\label{sec: terminal_velocity}

Figure~\ref{fig: vfinal_all} shows the predicted terminal velocity, $\varv_\infty$, and the terminal-to-escape velocity ratio, $\varv_{\infty}/\varv_{\text{esc},\Gamma}$, \footnote{The escape velocity is defined at the inner boundary radius and corrected for $\Gamma_\mathrm{e}$.} as a function of $T_\star$. The terminal velocity exhibits a clear temperature dependence, decreasing towards lower temperatures. 

In the hot regime ($T_\star \gtrsim 30\,\mathrm{kK}$), our models span velocities from approximately $500\,\mathrm{km\,s^{-1}}$ up to $4000\,\mathrm{km\,s^{-1}}$, depending on the wind strength. The colour-coding indicates that higher mass-loss rates generally correspond to lower terminal velocities at fixed temperature. Such an anticorrelation is consistent with the momentum equation: to first order, denser winds reduce radiative acceleration in the outer atmosphere, lowering the terminal velocities. 

But the entire velocity range might not correspond to single O-star configurations. Typical O-dwarfs have low wind strengths and they generally populate the higher end of the terminal velocity range shown here from roughly $2000-4000\,\mathrm{km\,s^{-1}}$. The lowest predicted terminal velocities, of order a few hundred $\mathrm{km,s^{-1}}$, may not be realised in single-star configurations. However, such velocities can be realised when accounting for binary interaction products, such as stripped stars which can have terminal velocities of order $\approx$$500-1000\,\mathrm{km\,s^{-1}}$~\citep[e.g.][]{Ramachandran2024, Muller_horn2026}. 

At cooler temperatures, the terminal velocities converge to roughly $500\,\mathrm{km\,s^{-1}}$. The ratio $\varv_{\infty}/\varv_{\text{esc},\Gamma}$ also decreases with decreasing $T_\star$, transitioning from a maximum of roughly $3$ at high temperatures to roughly $1-1.5$ at lower temperatures, broadly consistent with the empirical scalings of \citet{Lamers95}.

\subsection{VMSs as sources of ionising flux}
\label{sec: ionising flux}

In Fig.~\ref{fig: ionising_flux}, we show the predicted ionising photon luminosities for \ion{H}{I} ($\log Q_{\mathrm{HI}}$) and \ion{He}{II} ($\log Q_{\mathrm{HeII}}$) as a function of $T_\star$. Both quantities exhibit a strong temperature dependence, with several orders of magnitude drop across the tested temperature regime. The \ion{H}{I} ionising flux reaches a maximum of $\log Q_{\mathrm{HI}} \approx 50$ (in units of photons per second) at the hottest temperatures. As the temperature decreases, the \ion{H}{I} flux steadily decreases followed by a steep drop towards the coolest models in our grid. Similarly, $Q_{\mathrm{HeII}}$ also shows a steep drop towards cooler temperatures. This behaviour is physically expected: ionising photon production is governed by the high-frequency Wien tail of the spectrum, which shifts to lower energies as temperature decreases.

The colour-coding shows that at fixed temperature, higher mass-loss rates correspond to lower ionising fluxes. This suppression occurs because denser winds increase the optical depth in the UV, preventing ionising photons from escaping. This effect is particularly pronounced for \ion{He}{II}, which requires higher-energy photons that are more readily absorbed in the wind. A very similar effect is seen in classical WR winds where above a certain Eddington ratio, the ionising flux counts plummet \citep{Sander2020b}. The terminal velocities and ionising flux counts of all our models are listed in Table~\ref{tab: wind_properties}.

\begin{figure}
    \includegraphics[width = \columnwidth]{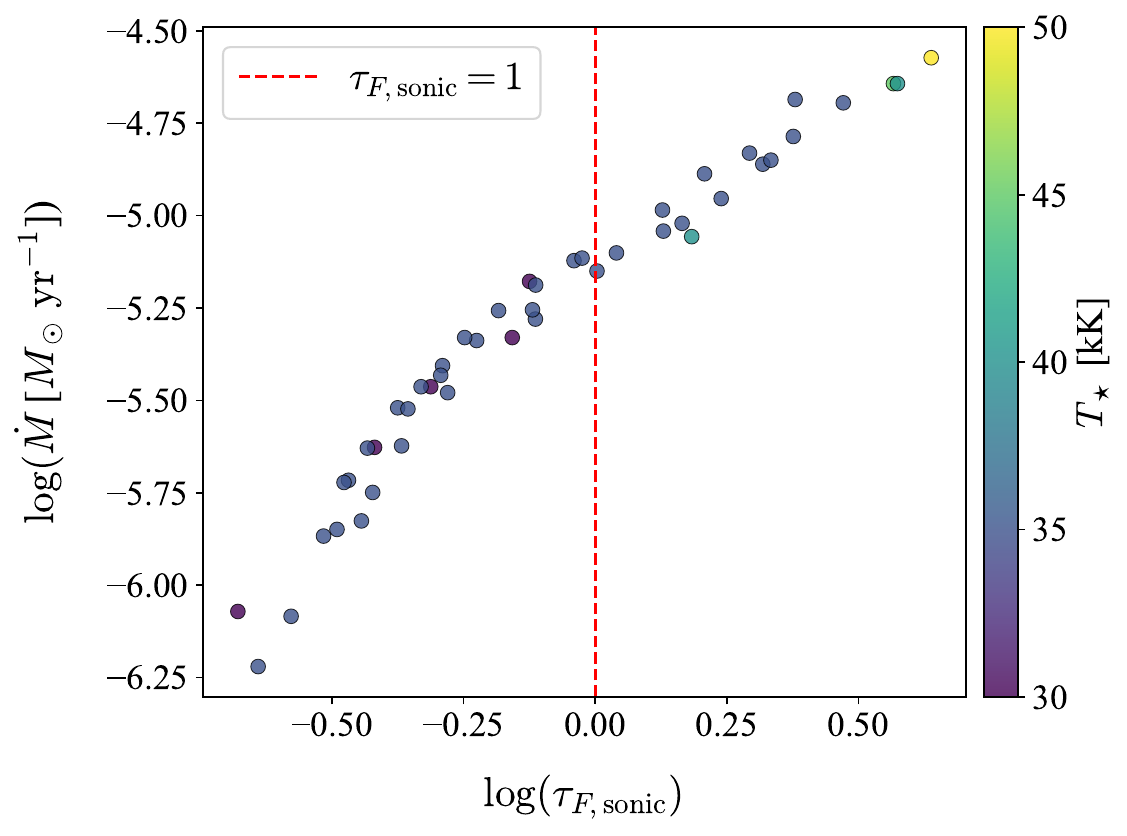}
    \caption{Predicted mass-loss rate as a function of the flux-weighted optical depth at the sonic point. All models with $5.7 < \log(L_\star/L_\odot) < 6.3$ and $25\,\mathrm{kK} < T_\mathrm{eff}(\tau_{R} = 2/3) < 35\,\mathrm{kK}$ are plotted regardless of surface H mass fraction, stellar mass (and therefore $\Gamma_\mathrm{e}$), or turbulent velocity. The dashed red line marks unity optical depth, separating optically thin from optically thick models. The symbols are colour-coded according to inner boundary temperature.}
    \label{fig: mdot_tau}
\end{figure}

\section{Discussion}
\label{sec: Discussion}

\subsection{Agreement with the transition mass-loss rate in the Arches Cluster}
\label{sec: Arches_compare}

A model-independent mass-loss point exists against which all mass-loss recipes can be tested. It is known as the `transition mass-loss point' and occurs at the spectral morphological transition between O-type and WNh stars. As discussed in Sect.~\ref{sec: kink}, a simple condition applies at this transition: $\eta \approx 0.6\tau_{F,\mathrm{sonic}}$, with $\tau_{F,\mathrm{sonic}}\approx1$. This enables its application to young clusters containing stars along the O--to--WNh sequence, including transitional Of/WNh objects.

The transition mass-loss rate is effectively a luminosity estimate expressed in units of mass loss. Since it depends primarily on the luminosity, it is significantly less model-dependent and more accurate than other mass-loss diagnostics. The transition mass-loss point therefore offers a unique opportunity to test mass-loss recipes: agreement at this point provides a first-order sanity check on the accuracy of a given recipe.

In the Arches Cluster, the transition occurs at a luminosity of approximately $\log(L_\star/L_\odot) \approx 6$, corresponding to a transition mass-loss rate of $\log(\dot{M}_{\mathrm{trans}}\,[M_\odot\,\mathrm{yr}^{-1}]) \approx -5.2$ \citep{Vink2012}. We can compare what our mass-loss prescription predicts at this optically thin-to-thick transition. The location of the kink in our models exactly matches $\tau_{F,\mathrm{sonic}}\approx 1$ and the synthetic spectra transition from absorption to emission (Paper~I and Sect.~\ref{sec: kink}). 

In Fig.~\ref{fig: mdot_tau}, we show our predicted mass-loss rate as a function of the wind optical depth $\tau_{F,\mathrm{sonic}}$ for all models within $5.7 < \log(L_\star/L_\odot) < 6.3$ and $25\,\mathrm{kK} < T_\mathrm{eff}(\tau_{\mathrm{R}} =2/3) < 35\,\mathrm{kK}$, corresponding to the relevant luminosities and temperatures covering the O--to--WNh transition in the Arches Cluster. No constraints were placed on the surface H mass fraction, the stellar mass (and therefore $\Gamma_\mathrm{e}$), or the turbulent velocity (see Appendix~\ref{appendix: vturb_fits} for the varying $\varv_\mathrm{turb}$ grid). We find an extremely tight relation between the mass-loss rate and the wind optical depth. Therefore, the mass-loss rate at $\tau_{F,\mathrm{sonic}} = 1$ has a robust value of $\log(\dot{M}_{\texttt{PoWR}^{\textsc{hd}}}[M_\odot\,\mathrm{yr^{-1}}]) \approx -5.15\pm0.1$, in excellent agreement with the transition mass-loss rate. 

\begin{figure}
    \includegraphics[width = \columnwidth]{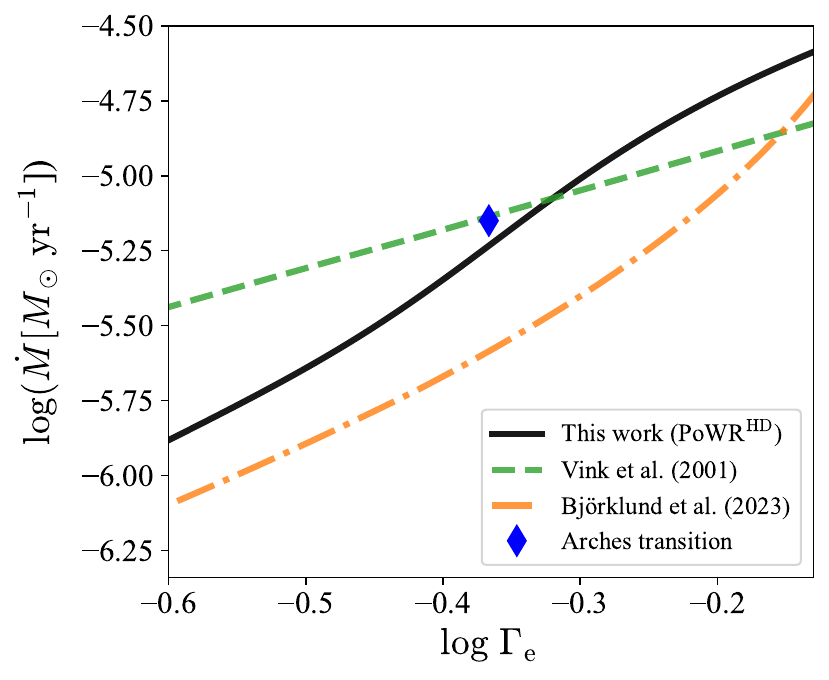}
    \caption{Comparison of our mass-loss predictions with oft-used massive star recipes as a function of $\Gamma_\mathrm{e}$. The luminosity and surface effective temperature are fixed to $\log(L_\star/L_\odot) = 6$ and $T_\mathrm{eff}(\tau_{\mathrm{R}} =2/3) = 35\,\mathrm{kK}$, respectively, corresponding to the Arches transition objects. The Arches transition mass-loss rate from \citet{Vink2012} is shown as a blue diamond, and its corresponding $\Gamma_\mathrm{e}$ is computed by adopting the stellar mass at which our hydrodynamic models predict a P Cygni morphology in H$\beta$ -- the characteristic spectroscopic signature of Of/WNh transition stars.}
    \label{fig: mdot_gamma_transition}
\end{figure}

In Fig.~\ref{fig: mdot_gamma_transition}, we further compare different mass-loss predictions against the transition mass-loss rate in the Arches Cluster. The stellar luminosity and surface effective temperature are fixed to $\log(L_\star/L_\odot) = 6$ and $T_\mathrm{eff}(\tau_{\mathrm{R}} =2/3) = 35\,\mathrm{kK}$, corresponding to the Arches transition objects, while the stellar mass is varied. Both the \citet{Vink2001} rates and our hydrodynamical predictions agree very well at the Arches Cluster O--to--WNh transition.

In summary, our mass-loss recipe is in excellent agreement with the transition mass-loss rate in the Arches Cluster. While several attempts have been made in recent years to capture the optically thin-to-thick transition using $\eta$ or $\Gamma_\mathrm{e}$ switches \citep{Vink2011,Chen2015,Graf2021,Sabhahit2023}, the novelty of our recipe lies in describing the complex mass-loss behaviour predicted by detailed hydrodynamical wind models with a single continuous function across the entire O--to--WNh parameter regime, while naturally capturing the physics of the single-scattering limit being crossed.

\begin{figure}
    \includegraphics[width = \columnwidth]{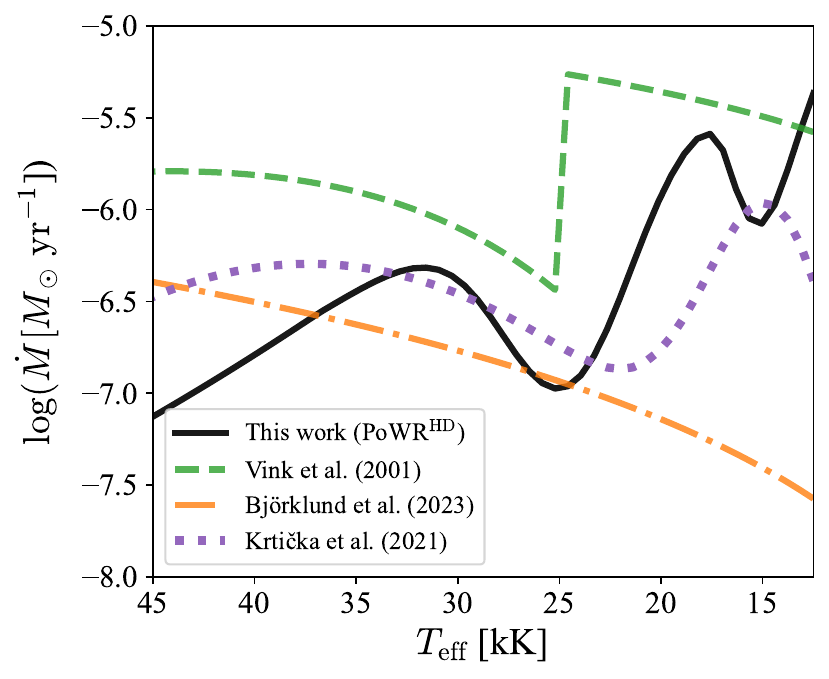}
    \caption{Comparison of our mass-loss predictions with oft-used massive-star recipes as a function of $T_\mathrm{eff}(\tau_\mathrm{R}=2/3)$. The stellar mass and luminosity are fixed to $M_\star = 50\,M_\odot$ and $\log(L_\star/L_\odot) = 5.6$, respectively.}
    \label{fig: mdot_Teff_compare}
\end{figure}

\subsection{Comparison to other recipes}
\label{sec: compare_recipe}

An accurate reproduction of the transition mass-loss rate, however, does not guarantee the accuracy of a recipe both above and below the transition region. For example, in Fig.~\ref{fig: mdot_gamma_transition}, we notice that while both the \citet{Vink2001} rates and our hydrodynamical predictions agree at the Arches transition, the two prescriptions exhibit very different behaviour above and below the transition. Below the transition, we consistently predict lower mass-loss rates than the \citet{Vink2001} rates, in better agreement with recent theoretical and empirical rates predicting a down-revision in rates of canonical 10-30 $M_\odot$ O stars \citep{VinkR, Bjorklund2023, Pauli2025}. 

Above the transition, however, our recipe surpasses the \citet{Vink2001} rates, in agreement with the \citet{Vink2011} findings. However, we note that the $\dot{M}$--$\Gamma_\mathrm{e}$ scaling above the transition point is not as steep as expected, which is simply due to non-linear temperature effects. Figure~\ref{fig: mdot_gamma_transition} uses a fixed $T_\mathrm{eff}(\tau_{\mathrm{R}} =2/3)$ criterion, while the shallow-to-steep scaling change in our models (cf. Fig.~\ref{fig: mdot_gammae_all}) was predicted for a fixed inner boundary $T_\star$. The solid black line in Fig.~\ref{fig: mdot_gamma_transition} shows the overall rates after accounting for the growing difference between the surface and inner boundary temperatures with denser winds (cf. Sect.~\ref{sec: fit_relations}).  

We also compare mass-loss rates as a function of $T_\mathrm{eff}(\tau_{\mathrm{R}} =2/3)$ for a fixed mass $M_\star = 50\,M_\odot$ and luminosity $\log(L_\star/L_\odot) = 5.6$. There is good qualitative agreement in the prediction of the first bistability jump between \citet{Vink2001}, \citet{Krticka21}, and our rates, albeit with the absolute \citet{Vink2001} rates being higher compared to ours, as discussed above. However, the \citet{Bjorklund2023} rates show no evidence of a bistable behaviour~\citep[see][for a recent overview and robustness of the bistability predictions]{Vink2026}. 

\subsection{Application of our rates to the zero-age main sequence}
\label{sec: ZAMS_application}

\begin{table*}[h]
\centering
\caption{Mass-loss predictions on the ZAMS.}
\label{tab: zams_selected}
\begin{tabular}{cccccccc}
\hline
$\log (L_\star/L_\odot)$ & $T_\mathrm{eff}(\tau_\mathrm{R}=2/3)$ [kK] & $M_\star/M_\odot$ & $X$ & $\Gamma_\mathrm{e,high\text{-}T}$ & $T_\star$ & $\Gamma_{\mathrm{e}}$ & $\log (\dot{M}_{\texttt{PoWR}^{\textsc{hd}}}[M_\odot\,\mathrm{yr^{-1}}])$ \\
\hline
5.084 & 40.006 & 30 & 0.7 & 0.106 & 41.012 & 0.106 & $-8.370$ \\
5.365 & 43.376 & 40 & 0.7 & 0.151 & 44.398 & 0.151 & $-7.693$ \\
5.566 & 45.602 & 50 & 0.7 & 0.193 & 46.662 & 0.193 & $-7.189$ \\
5.721 & 47.094 & 60 & 0.7 & 0.229 & 48.219 & 0.229 & $-6.807$ \\
5.950 & 48.665 & 80 & 0.7 & 0.291 & 49.995 & 0.291 & $-6.273$ \\
6.115 & 49.004 & 100 & 0.7 & 0.341 & 50.616 & 0.341 & $-5.909$ \\
6.244 & 48.550 & 120 & 0.7 & 0.382 & 50.495 & 0.382 & $-5.636$ \\
6.395 & 47.168 & 150 & 0.7 & 0.432 & 49.640 & 0.432 & $-5.332$ \\
6.512 & 45.412 & 180 & 0.7 & 0.473 & 48.420 & 0.473 & $-5.099$ \\
6.579 & 44.214 & 200 & 0.7 & 0.495 & 47.589 & 0.495 & $-4.967$ \\
6.715 & 41.425 & 250 & 0.7 & 0.542 & 45.815 & 0.542 & $-4.681$ \\
6.822 & 38.967 & 300 & 0.7 & 0.579 & 44.565 & 0.579 & $-4.436$ \\
6.986 & 35.533 & 400 & 0.7 & 0.633 & 43.725 & 0.633 & $-4.062$ \\
\hline
\end{tabular}
\end{table*}

We finally applied our rates using more realistic stellar $L_\star$ and $T_\mathrm{eff}(\tau_{\mathrm{R}} =2/3)$ along the zero-age main sequence (ZAMS). Table~\ref{tab: zams_selected} summarises the input mass, H mass fraction, surface luminosity, and surface temperature for models on the ZAMS. The luminosity and surface temperature for each input mass are obtained from the Modules for Experiments in Stellar Astrophysics (MESA) stellar structure and evolution code (see Appendix~\ref{appendix: MESA_structure} for code specifics and inputs) when H begins to burn in the core (i.e. when $X_\mathrm{core}(t) < X_\mathrm{core}(t=0) - 0.01$).

Among the ZAMS models, we see a clear pattern in the predicted mass-loss rate and the resulting $T_\star$. As the initial mass increases, the predicted mass-loss rate likewise increases and the difference between $T_\star$ and $T_\mathrm{eff}(\tau_\mathrm{R}=2/3)$ grows, successfully capturing the behaviour seen in our hydro-models. We also notice that on the ZAMS, the temperature is hot enough for the high-temperature approximation of $\Gamma_\mathrm{e}$ to hold.

\begin{figure}
    \includegraphics[width = \columnwidth]{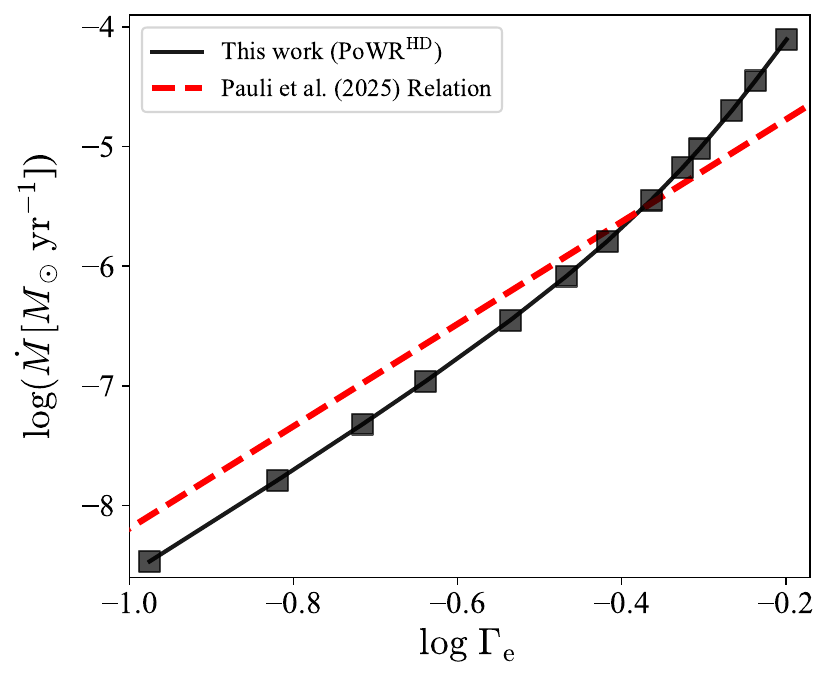}
    \caption{Mass-loss prediction on the ZAMS as a function of the Eddington parameter, $\Gamma_\mathrm{e}$, (solid black line with grey squares). The regression relation, derived from the \citet{Pauli2025} relation based on empirical masses, is shown as a dashed red line.}
    \label{fig: ZAMS}
\end{figure}

We show the full sequence of our $30-400\,M_\odot$ ZAMS models in Fig.~\ref{fig: ZAMS}. The kink in the mass-loss rate is not as prominent, as one would expect from a power-law slope switch from $\sim$$2$ to $\sim$$10$. This is mainly due to a combination of the non-linear temperature effects mentioned previously and the location of the kink itself gradually shifting to higher $\Gamma_\mathrm{e}$ as the luminosity increases, which effectively blurs the kink in $\Gamma_\mathrm{e}$ space. 

Ideally, we would compare our results with empirical $\dot{M}$--$\Gamma_\mathrm{e}$ scalings that use dynamical masses to determine Eddington parameters, which are more accurate than spectroscopic masses. However, this would severely limit the sample size for a meaningful comparison. We therefore adopted the sample compilation from \citet{Pauli2025}, who selected stars with UV spectra and empirically determined masses that are approximately equal to their evolutionary masses, or dynamical masses where available. A slight bend occurs in our predicted $\dot{M}$--$\Gamma_\mathrm{e}$ scaling on the ZAMS compared with the empirical relation but only above $M_\mathrm{ZAMS} > 250\,M_\odot$, which becomes statistically difficult to detect given the scarcity of VMSs and the unavailability of dynamical mass estimates for such high masses in the current empirical data. With evolution, however, it might become easier to detect the kink at lower initial masses as stars naturally evolve to higher $\Gamma_\mathrm{e}$, although this will depend on the accuracy of both $\dot{M}$ and $\Gamma_\mathrm{e}$ determinations. We compare the mass-loss rate only on the ZAMS in the current work. A full integration into an evolutionary code with detailed evolutionary models across a range of masses will be presented in forthcoming work.

\section{Conclusions}
\label{sec: conclusions}

We presented a new mass-loss prescription for massive stars and VMSs based on a comprehensive grid of hydrodynamically consistent \texttt{PoWR}$^{\textsc{hd}}$ wind-atmosphere models. Our study spans a wide range of stellar luminosities, masses, surface He over H enhancement, temperatures, and Eddington parameters, allowing us to capture the complex $\Gamma$ and temperature dependence of radiatively driven mass loss across the O star to WNh regime.

Our main results can be summarised as follows:

\begin{enumerate}

\item We confirm the presence of a wind mass-loss kink in the $\dot{M}$--$\Gamma_\mathrm{e}$ relation across the explored parameter space. The kink marks a transition from a shallow scaling ($\sim$$2$) at low $\Gamma_\mathrm{e}$ to a steeper scaling ($\sim$$10$) at high $\Gamma_\mathrm{e}$. 
We captured the kink with a smooth log-sum-exponential formulation that naturally connects the two different slope regimes. Whilst $\Gamma_\mathrm{e}$ is the primary controlling parameter, we demonstrated that additional explicit dependencies on luminosity and surface H mass fraction are required to fully reproduce the model results. A single-parameter $\Gamma_\mathrm{e}$ scaling is insufficient.

\item We showed that the kink coincides with a narrow range of wind optical depths and wind efficiency parameters, with $\tau_{F,\mathrm{sonic}} \approx 1$ and $\eta \approx 0.4$. This confirms that the kink is fundamentally linked to the onset of optically thick winds and the transition from single to multiple scattering. This makes the present prescription  qualitatively consistent with the $\eta$--framework of \citet{Sabhahit2022}, which implemented the mass-loss kink predicted from MC models and anchored the absolute rate at the kink to the Arches Cluster transition mass-loss rate using the $\eta$ criterion. 

\item Our prescription reproduces the model-independent transition mass-loss rate inferred for the Arches Cluster. At the observed transition luminosity, we predict $\log(\dot{M}_{\texttt{PoWR}^{\textsc{hd}}}[M_\odot\,\mathrm{yr^{-1}}]) \approx -5.15$, in excellent agreement with the transition mass-loss rate of $\log(\dot{M}_{\mathrm{trans}}\,[M_\odot\,\mathrm{yr}^{-1}]) \approx -5.2$, providing a robust sanity check on our recipe.

\item Below the transition, our predicted mass-loss rates are systematically lower than the oft-used \citet{Vink2001} rates, in better agreement with many empirical results. Above the transition, our rates increase with $\Gamma_\mathrm{e}$ and qualitatively agree with dynamically consistent MC predictions from \citet{Vink2011}.

\item The temperature dependence of mass loss is governed by two distinct effects: a general increasing mass-loss trend towards larger radii and superposed bistability jumps associated with changes in the iron ionisation balance. We identified two bistability jumps and incorporated them into the final fitting relations.

\item We provided predictions for terminal velocities and ionising photon luminosities across our parameter space. Terminal velocities range roughly from $500-4000\,\mathrm{km\,s^{-1}}$ depending on temperature and wind strength, with $\varv_{\infty}/\varv_{\text{esc},\Gamma}$ ratios decreasing from roughly $3$ at high temperatures to roughly $1-1.5$ at lower temperatures. Ionising fluxes show strong temperature dependence and are suppressed at fixed temperature by increasing mass-loss rates due to wind opacity effects.

\item For practical implementation in stellar evolution models, we provided auxiliary relations linking the inner-boundary temperature ($T_\star$) to the effective temperature at $\tau_\mathrm{R}=2/3$, as well as a temperature-dependent correction to the classical Eddington parameter accounting for incomplete ionisation at lower temperatures.

\end{enumerate}

Our results demonstrate that mass loss from massive stars is governed by a combination of Eddington proximity, luminosity, temperature, and H abundance through its influence on the electron density, rather than by any single parameter alone. The prescription presented here is based on hydrodynamically consistent wind calculations and shows excellent agreement with both the O star to WNh transition rate in the Arches Cluster and the recent $\dot{M}$--$\Gamma_\mathrm{e}$ relation over a wide range of $\Gamma_\mathrm{e}$. This makes it well-suited for stellar evolution calculations of massive stars and VMSs in the $20-500\,M_\odot$ range. Future work will focus on coupling this prescription self-consistently into evolutionary models, exploring its impact on the evolution of VMSs, as well as an extension of the hydro-model grid to low $Z$.

\section*{Data availability}

The complete grid of hydrodynamically consistent $\texttt{PoWR}^{\textsc{HD}}$ model predictions listed in Table~\ref{tab: wind_properties} is only available in electronic form at the CDS via anonymous ftp to cdsarc.u-strasbg.fr (130.79.128.5) or via http://cdsweb.u-strasbg.fr/cgi-bin/qcat?J/A+A/.

\begin{acknowledgements}
We thank the anonymous referee for constructive comments that helped improve the paper.
GNS and JSV are supported by STFC funding under grant number ST/Y001338/1. AACS is supported by the German
    \emph{Deut\-sche For\-schungs\-ge\-mein\-schaft, DFG\/} in the form of an Emmy Noether Research Group -- Project-ID 445674056 (SA4064/1-1, PI Sander). AACS further acknowledges financial support by the Federal Ministry of Research, Technology and Space (BMFTR) via the German Aerospace Center (Deutsches Zentrum f\"ur Luft- und Raumfahrt, DLR) grant 50 OR 2509 (PI: Sander). This project was co-funded by the European Union (Project 101183150 - OCEANS).
\end{acknowledgements}

\bibliographystyle{aa}
\bibliography{References.bib}

\begin{appendix}

\section{Turbulent velocity correction to the mass-loss rate}
\label{appendix: vturb_fits}

The mass-loss recipe in Eq.~\eqref{eq: mdot_full_fit} is based on atmosphere models with a radially constant turbulent velocity fixed at $\varv_\mathrm{turb} = 70.71\,\mathrm{km\,s^{-1}}$. However, multi-D simulations find large turbulent velocities of order $\sim$$10-200\,\mathrm{km\,s^{-1}}$ originating from radiatively driven turbulence increasing outwards from the iron opacity bump. \citet{Moens2025} recently found the maximum turbulent velocities reached in the atmosphere to be $\Gamma$--dependent. Implementation of a multi-D informed prescription for depth-dependent turbulence, which would be required to consistently model the entire envelope structure from the iron opacity bump to the supersonic wind, is currently underway in the $\texttt{PoWR}^\textsc{hd}$ models. For now, we note that the inner boundary in our simulations has a fixed Rosseland continuum optical depth of 20, placing it well above the iron opacity bump and in the regime where significant turbulence is already realised in the multi-D models, and provide a mass-loss correction term for different values of radially constant $\varv_\mathrm{turb}$ in our atmosphere.

\begin{figure}
    \includegraphics[width = \columnwidth]{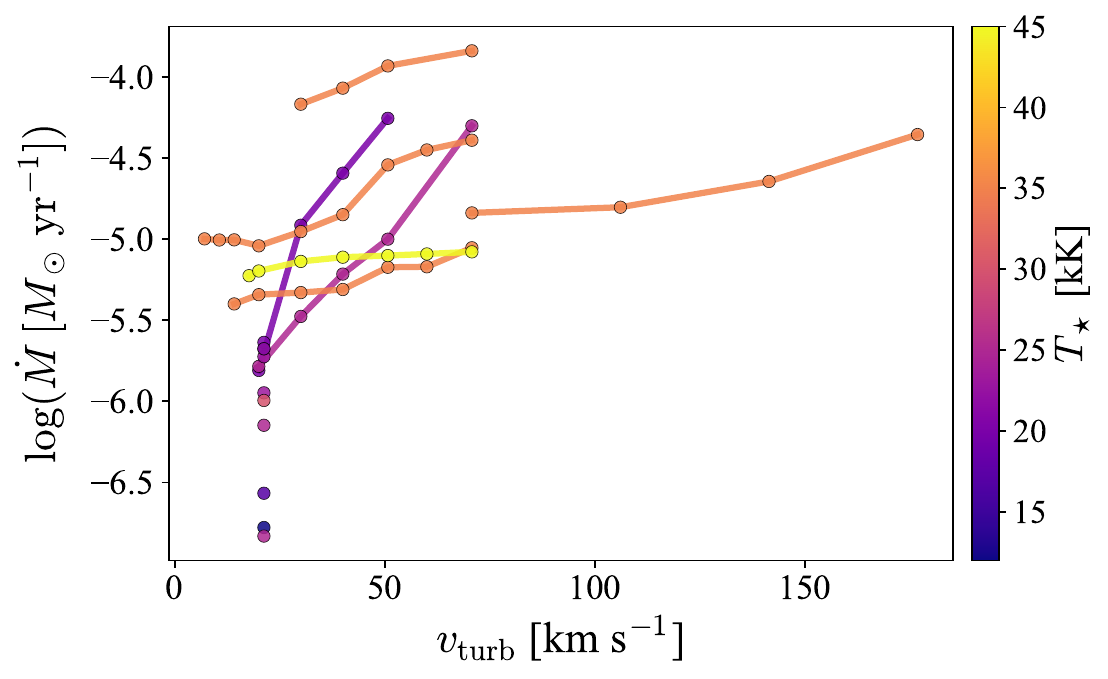}
    \caption{Predicted mass-loss rates as a function of $\varv_\mathrm{turb}$, colour-coded by the inner boundary temperature, $T_\star$. The symbols connected by solid lines represent model sequences in which all input stellar parameters are fixed except $\varv_\mathrm{turb}$. }
    \label{fig: mdot_vturb}
\end{figure}

The physical basis for the correction is that increasing $\varv_\mathrm{turb}$ adds additional pressure gradient support in the hydrodynamic equation, that is, the $\Gamma$ term associated with turbulence, $\Gamma_\mathrm{turb}$, increases. For the majority of cases, increasing $\varv_\mathrm{turb}$ results in a higher mass-loss rate, as seen in Fig.~\ref{fig: mdot_vturb}.  However, a small subset of models (three out of the 44 models tested) show the opposite trend, where the mass-loss rate decreases with increasing $\varv_\mathrm{turb}$. We note that this opposite trend is confined to the lowest values of $\varv_\mathrm{turb}$ explored in our grid (of $\approx7-20\,\mathrm{km\,s^{-1}}$), which may be at the low end for VMSs. Figure~\ref{fig: mdot_vturb} also shows how $T_\star$ impacts the effect of $\varv_\mathrm{turb}$ on the mass-loss rate, with the steepest $\dot{M}$--$\varv_\mathrm{turb}$ slopes occurring at the lowest temperatures.

The correction is therefore applied directly to the $\Gamma_\mathrm{e}$ term such that a higher $\varv_\mathrm{turb}$ results in an upward correction to $\Gamma_\mathrm{e}$, which consequently increases the mass-loss rate. All $\Gamma_\mathrm{e}$ terms in Eq.~\eqref{eq: mdot_full_fit} are therefore replaced by a corrected term, $\Gamma_\mathrm{e,corr}$, of the form
\begin{equation}
\begin{split}
\Gamma_\mathrm{e,corr} = \Gamma_\mathrm{e}\cdot\Big(1+\alpha(\Gamma_\mathrm{e}, T_\star)\cdot(\varv_\mathrm{turb}/70.71-1)\Big),
\label{eq: vturb_correction_gamma_e}
\end{split}
\end{equation}
where $\alpha$ is obtained from a fit to the models from the varying $\varv_\mathrm{turb}$ grid as follows:
\begin{equation}
\begin{split}
\alpha(T_\star, \Gamma_\mathrm{e}) &= \alpha_{\min} + \alpha_{\mathrm{diff}} \cdot \frac{\Gamma_\mathrm{e}^{m}}{\Gamma_\mathrm{e}^{m} + \left(\dfrac{T_\star}{T_0}\right)^{n}} \\[6pt]
\alpha_{\min} &= 0.062 \err{0.083}, \quad \alpha_{\mathrm{diff}} = 0.713 \err{0.118} \\
T_0      &= 47.808 \err{2.851} \; \mathrm{kK} \\
m        &= 4.186 \err{1.416}, \quad n = 6.332 \err{1.952}
\end{split}
\label{eq:vturb_correction_alpha}
\end{equation}
where $T_\star$ is in kK. The typical error in the mass-loss rate predictions using this correction is 0.2 dex. The value of $\Gamma_\mathrm{e,corr}$ obtained in this way is clipped between 0.01 and 0.99 before it replaces the $\Gamma_\mathrm{e}$ terms in Eq.~\eqref{eq: mdot_full_fit}.

\section{MESA structure models for the ZAMS}
\label{appendix: MESA_structure}

In Sect.~\ref{sec: ZAMS_application}, we plotted mass-loss rates as a function of $\Gamma_\mathrm{e}$ for stars on the ZAMS using luminosities and surface temperatures obtained from MESA structure models. Here we detail the MESA inputs and configuration used to build our ZAMS models. 

We used MESA version r12115 to build our ZAMS grid. The initial masses are $30$, $40$, $50$, $60$, $80$, $100$, $120$, $150$, $180$, $200$, $250$, $300$, and $400\,M_\odot$. The H, He, and metal mass fractions are set to $X=0.7$, $Y=0.28$, and $Z=0.02$. The individual mass fractions of metals follow the solar-scaled abundances from \citet{GS98}. The models are non-rotating and have no mass loss. Convection is treated using the Ledoux criterion with the mixing-length theory (MLT) free parameter $\alpha_\mathrm{MLT}=1.5$. 

Most of the inputs have negligible impact on the ZAMS luminosities, but the ZAMS surface temperatures, especially for the highest-mass models that are close to the Eddington limit, can be affected by the choice of mixing parameters. For example, the ZAMS bends due to inflated layers above $\sim$$100\,M_\odot$, with the bending increasing towards higher masses. The models presented here use no modification of the temperature gradient in the radiatively dominated sub-surface convective layers as predicted by MLT; that is, we do not use MLT++ to compute the ZAMS.

\section{Parameter space and predicted wind properties}
\label{appendix: summary}

\onecolumn
\begin{landscape}
\centering
\small
\begin{longtable}{c c c c c | c c c c c c c c c c c c c}
\caption{Predicted wind properties from  $\texttt{PoWR}^\textsc{hd}$ models for input stellar parameters.}
\label{tab: wind_properties} \\
\hline
\rule{0pt}{14pt}
$\log\bigg(\dfrac{L_\star}{L_\odot}\bigg)$ & $\dfrac{M_\star}{M_\odot}$ & $X$ & $T_\star^\dag$ & $\varv_\mathrm{turb}$$^{\triangle}$ & $Y$ & $T_{\rm eff}(\tau_\mathrm{R}=2/3)^\dag$ & $\log(\dot{M})^\blacktriangle$ & $\varv_\infty$$^{\triangle}$ & $\eta$ & $\tau_{F}(r_\mathrm{sonic})$ & $\Gamma_\mathrm{e}$ & $\dfrac{R_\star}{R_{\odot}}$ & $\dfrac{R_{\rm crit}}{R_{\odot}}$ & $\dfrac{R(\tau_\mathrm{R} = 2/3)}{R_\odot}$ & $\dfrac{\varv_\infty}{\varv_{\rm esc,\Gamma}(R_\star)}$ & $\log(Q_\ion{H}{I})^\star$ & $\log(Q_\ion{He}{II})^\star$ \\[7pt]
\hline
\endfirsthead
\hline
\rule{0pt}{14pt}
$\log\bigg(\dfrac{L_\star}{L_\odot}\bigg)$ & $\dfrac{M_\star}{M_\odot}$ & $X$ & $T_\star^\dag$ & $\varv_\mathrm{turb}$$^{\triangle}$ & $Y$ & $T_{\rm eff}(\tau_\mathrm{R}=2/3)^\dag$ & $\log(\dot{M})^\blacktriangle$ & $\varv_\infty$$^{\triangle}$ & $\eta$ & $\tau_{F}(r_\mathrm{sonic})$ & $\Gamma_\mathrm{e}$ & $\dfrac{R_\star}{R_{\odot}}$ & $\dfrac{R_{\rm crit}}{R_{\odot}}$ & $\dfrac{R(\tau_\mathrm{R} = 2/3)}{R_\odot}$ & $\dfrac{\varv_\infty}{\varv_{\rm esc,\Gamma}(R_\star)}$ & $\log(Q_\ion{H}{I})^\star$ & $\log(Q_\ion{He}{II})^\star$ \\[7pt]
\hline
\endhead
\hline
\endfoot
\hline
\endlastfoot
5.50 & 16 & 0.70 & 35 & 70.7 & 0.28 & 13.60 & $-4.245$ & 469.36 & 4.15 & 15.44 & 0.52 & 15.34 & 25.43 & 101.36 & 1.07 & 0.00 & 28.56 \\
5.50 & 17 & 0.70 & 35 & 70.7 & 0.28 & 14.94 & $-4.444$ & 513.29 & 2.87 & 10.80 & 0.48 & 15.34 & 30.30 & 82.39 & 1.10 & 47.38 & 31.66 \\
5.50 & 20 & 0.70 & 40 & 70.7 & 0.28 & 30.27 & $-5.094$ & 737.65 & 0.92 & 3.13 & 0.41 & 11.74 & 19.10 & 20.46 & 1.20 & 49.16 & 36.37 \\
5.50 & 20 & 0.70 & 35 & 70.7 & 0.28 & 24.03 & $-5.054$ & 534.31 & 0.73 & 2.56 & 0.41 & 15.34 & 33.82 & 32.49 & 0.98 & 48.94 & 34.25 \\
5.50 & 20 & 0.70 & 30 & 70.7 & 0.28 & 17.35 & $-4.746$ & 550.67 & 1.54 & 5.70 & 0.39 & 20.87 & 69.22 & 61.75 & 1.17 & 47.77 & 30.13 \\
5.50 & 20 & 0.70 & 25 & 70.7 & 0.28 & 12.56 & $-4.302$ & 400.51 & 3.11 & 7.45 & 0.39 & 30.06 & 88.91 & 118.64 & 1.02 & 38.57 & 19.21 \\
5.50 & 20 & 0.70 & 23 & 70.7 & 0.28 & 8.31 & $-3.960$ & 379.64 & 6.47 & 28.23 & 0.39 & 35.51 & 92.19 & 267.00 & 1.05 & 37.93 & 20.32 \\
5.50 & 23 & 0.70 & 35 & 70.7 & 0.28 & 28.81 & $-5.364$ & 706.99 & 0.48 & 1.35 & 0.35 & 15.34 & 28.40 & 22.50 & 1.16 & 49.04 & 36.02 \\
5.50 & 25 & 0.70 & 35 & 70.7 & 0.28 & 30.09 & $-5.576$ & 927.44 & 0.38 & 0.88 & 0.33 & 15.34 & 27.13 & 20.67 & 1.43 & 49.11 & 37.03 \\
5.50 & 30 & 0.70 & 35 & 70.7 & 0.28 & 32.12 & $-5.902$ & 1101.60 & 0.21 & 0.46 & 0.27 & 15.34 & 23.10 & 18.19 & 1.49 & 49.11 & 37.75 \\
5.50 & 35 & 0.70 & 35 & 70.7 & 0.28 & 32.84 & $-6.165$ & 1381.57 & 0.15 & 0.30 & 0.23 & 15.34 & 22.08 & 17.41 & 1.69 & 49.14 & 38.11 \\
5.50 & 40 & 0.70 & 40 & 70.7 & 0.28 & 38.69 & $-7.024$ & 5803.87 & 0.09 & 0.12 & 0.21 & 11.74 & 16.67 & 12.55 & 5.73 & 49.25 & 42.51 \\
\end{longtable}
\tablefoot{This table is available in its entirety at the CDS. A portion is shown here for guidance regarding its form and content.\\
$^\dag$ Temperatures, $T_\star$ and $T_{\rm eff}(\tau_\mathrm{R}=2/3)$, are expressed in kilokelvin.\\
$^\triangle$ Turbulent velocity $\varv_\mathrm{turb}$ and terminal velocity $\varv_{\infty}$ are expressed in kilometres per second.\\
$^\blacktriangle$ Mass-loss rate $\dot{M}$ is expressed in solar masses per year.\\
$^\star$ $Q_\ion{H}{I}$ and $Q_\ion{He}{II}$ are expressed in number of photons per second.}
\end{landscape}

\end{appendix}

\end{document}